\documentclass{aa} 
\usepackage{txfonts}
\usepackage{graphicx}
\usepackage{appendix}
\usepackage{natbib}
\usepackage{url}
\usepackage{subfigure}
\bibpunct{(}{)}{;}{a}{}{,}

\begin{document}

\title{On the absorption of X-ray bright broad absorption line quasars\thanks{Based on observations obtained with {XMM-{\textit{Newton}}}, an ESA science mission with instruments and contributions directly funded by ESA Member States and NASA.}}

\author{
	M.~Giustini\inst{1,}\inst{2},
	M.~Cappi\inst{2},
	\and
	C.~Vignali\inst{1,}\inst{3}
	}

\offprints{M.~Giustini, \email{giustini@iasfbo.inaf.it}}

\institute{\inst{1}Dipartimento di Astronomia, Universit\`a degli Studi di Bologna, via Ranzani 1, I-40127, Bologna, Italy\\
           \inst{2}Istituto di Astrofisica Spaziale e Fisica Cosmica (INAF), via Gobetti 101, I-40129, Bologna, Italy \\
           \inst{3}Osservatorio Astronomico di Bologna (INAF), via Ranzani 1, I-40127, Bologna, Italy}

\date{Received XXXXXXXXXX XX, XXXX; accepted XXXXXXXXXX XX, XXXX}

\abstract
 {
 	Most X-ray studies of broad absorption line quasars (BALQSOs) found significant ($N_{\rm{H}}\sim 10^{22-24}$~cm$^{-2}$) intrinsic column densities of gas absorbing an underlying typical power-law continuum emission, in agreement with expectations from radiatively driven accretion disk wind models. However, direct spectral analysis was performed only on a limited number of bright sources.}
 {
 	We investigate the X-ray emission of a large number of BALQSOs at medium to high redshift ($0.8\lesssim z \lesssim 3.7$) with the best data available to date.}
 {
	We drew a large BALQSO sample from the cross-correlation of SDSS DR5 and 2XMM catalogs to perform moderate-quality X-ray spectral and hardness ratio analysis and X-ray/optical photometry.
 We compare our results with previous studies of BALQSOs and theoretical disk wind model expectations.}
 {
 	No or little intrinsic X-ray neutral absorption is found for one third of the spectroscopically analyzed BALQSO sample ($N_{\rm{H}} < 4 \times 10^{21}\:$cm$^{-2}$ at 90\% confidence level), and lower than typical X-ray absorption is found in the remaining sources ($\langle N_{\rm{H}}\rangle \sim 5\times 10^{22}\:$cm$^{-2}$) even including the faintest sources analyzed through hardness ratio analysis. The mean photon index is $\Gamma \sim 1.9$, with no significant evolution with redshift. The optical/X-ray spectral indices $\alpha_{\rm{ox}}$ are typical of radio-quiet broad line AGN, in contrast with the known (from previous X-ray studies) ``soft X-ray weakness'' of BALQSOs and in agreement with the lack of X-ray absorption. We found the low-absorption index (AI) subsample to host the lowest X-ray absorbing column densities of the entire sample. }
 {
	X-ray selected BALQSOs show lower X-ray absorption than purely optically selected ones, and soft X-ray weakness does not hold for any of them. Their outflows may be launched by different mechanisms than classical soft X-ray weak BALQSOs or they may be the tail of the already known population seen along a different line of sight, in both cases expanding the observational parameter space for their search and investigation.}

\keywords{X-rays:general - galaxies:active - quasars:absorption lines - quasars:general }

\titlerunning{On the absorption of X-ray bright BALQSOs}
\authorrunning{M.~Giustini et al.}

\maketitle


\section{Introduction}\label{sect:intro}

Recent observational works have unveiled the presence of outflows in many radio-quiet active galactic nuclei (AGN). Such outflows may be relevant both for the unification schemes and modelization of AGN \citep[e.g.][]{2000ApJ...545...63E} and for the growth of cosmic structures, partially providing the feedback needed to reconcile observations with theoretical expectations \citep[e.g.][]{2005Natur.433..604D,2005ApJ...635L..13S,2005ApJ...619...60L}.

We know that AGN outflows are there and are ionized, since we observe them as blueshifted absorption features in their UV/X-rays continua \citep[warm absorbers, ultraviolet broad absorption lines, and X-ray absorption lines, see e.g.][]{2003ARA&A..41..117C, 2006AN....327.1012C, 2007ApJ...659.1022K}.
We still have a limited knowledge of the physics responsible for launching and accelerating the outflows/wind.
The driver of these winds may be either radiation, thermal, or magnetic pressure \citep[e.g.][]{2007ASPC..373..267P}. Their launching region may lay between a few hundreds gravitational radii \citep[as estimated by variability studies on blueshifted X-ray absorption lines, e.g.][]{2007ApJ...670..978B} and a few parsecs \citep[where thermal evaporation of outer dusty torus takes place, e.g.][]{2001ApJ...561..684K}. 
The wind may be equatorial, as in accretion disk winds \citep{1995ApJ...451..498M, 2000ApJ...543..686P}, and/or polar \citep{1999ApJ...527..624P,2007ApJ...661..693P}. The wind may be continuous or made of knots/blobs of matter \citep[as in the model of][]{2004A&A...413..535G}. 

The vast majority of both theoretical and observational works about AGN outflows concerns the first class of objects discovered to host strong nuclear outflows: the broad absorption line quasars (BALQSOs). 
These are $10-15\%$ of optically selected QSOs\footnote{The fraction is higher, $\gtrsim 20\%$, for near-infrared and radio selected QSOs \citep{2008ApJ...672..108D, 2008arXiv0801.4379S}, so maybe suggesting a bias in optical band against BALQSOs detection.} that show broad ($\gtrsim 2000\:$km~s$^{-1}$) UV resonant absorption lines strongly blueshifted with respect to the systemic velocities of the source, indicating outward motions along the line of sight with typical velocities of several $10^3\:$km~s$^{-1}$ and extending up to $\sim 60\,000\:$km~s$^{-1}$ \citep[e.g. Q1414+087,][]{1983PASP...95..341F}. 
The observed fraction of BALQSOs could correspond either to the covering fraction of a wind present in all AGN or to the intrinsic fraction of QSOs hosting (totally covering) massive nuclear winds, so tracing an evolutionary phase of the AGN activity lasting $\sim 10-15\%$ of their lives. 
The great similarites between the optical/UV emission line and continuum properties of BALQSOs and non-BALQSOs \citep[e.g.][]{2003AJ....126.2594R} support the first scenario.
In the latter scenario, the large amount of gas and dust surrounding the central source should lead to enhanced far-infrared and submillimeter emission in BALQSOs with respect to non-BALQSOs. Recent studies have found no \citep{2003ApJ...598..909W} or little \citep{2007MNRAS.374..867P} differences among the two populations; however deeper observations of larger samples are needed to assess a firm conclusion about this point.

X-ray studies can help to constrain the physical mechanisms responsible for launching and accelerating AGN winds, and discriminate among different scenarios for BALQSOs.
In the popular model of radiatively driven accretion disk wind, the high column densities of X-ray absorbing gas -- postulated in the original form by \citet{1995ApJ...451..498M} in order to prevent overionization of the wind by the strong central continuum -- arise naturally in the hydrodynamical 2D simulations of \citet{2000ApJ...543..686P} and \citet{2004ApJ...616..688P}. If accretion disk wind models hold, X-ray observations of BALQSO should thus reveal high ($N_{\rm{H}}>10^{23}\:$cm$^{-2}$) columns of gas absorbing the primary continuum and shielding the UV wind.

Since the time of ROSAT observations, BALQSOs are known to be rather dim in X-rays \citep{1995ApJ...450...51G}. Their X-ray flux, typically $10-30$ times lower than expected from their UV flux, often qualifies them as 'soft X-ray weak' objects \citep{1997ApJ...477...93L}. The discovery of a correlation between C~IV equivalent width and soft X-ray weakness \citep{2000ApJ...528..637B} strongly suggests large absorption to be the reason for the low X-ray flux of BALQSO, so supporting the accretion disk wind models and the underlying typical QSO spectral energy distribution (SED).

A two-fold strategy has been pursued in the last years in order to overcome the X-ray weakness of BALQSOs and determine their high-energy properties. A hardness ratio and/or stacking analysis technique was applied to snapshot observations of a large number of sources in order to characterize their mean properties \citep[e.g.][]{2001ApJ...558..109G, 2006ApJ...644..709G}, while detailed spectral analysis of deep observations was only possible for a few bright sources \citep[e.g.][]{2002ApJ...567...37G,2003AJ....126.1159G, 2004ApJ...603..425G, 2005AJ....130.2522S, 2005A&A...433..455S}. Both kind of studies revealed high column densities ($N_{\rm{H}} \sim 10^{22-24}\:$cm$^{-2}$) of gas absorbing a 'normal' intrinsic SED with powerlaw spectral indices typical of radio-quiet, non-absorbed QSOs \citep[$\Gamma \sim 1.7-2$,][]{2005A&A...432...15P}. In almost none of the sources a simple neutral absorption scenario is able to well reproduce the data, but rather complex absorbers (ionized and/or partially covering the source and/or variables) are generally observed or deduced from the available data. To our knowledge, the only two solid cases of Compton-thick absorption (i.e. $N_{\rm{H}}\gtrsim 1.5\times 10^{24}\:$cm$^{-2}$) in BALQSOs have been found in the mini-BALQSO\footnote{mini-BALQSOs show somewhat narrower and less blueshifted BALs than BALQSOs. They form a continuous sequence in UV outflow properties together with Narrow Absorption Line (NAL) QSOs and BALQSOs, i.e. NAL $\rightarrow$ mini-BAL $\rightarrow$ BAL with increasing absorption troughs widths \citep[see e.g.][]{2004ASPC..311..203H}.} Mrk~231 \citep{2004A&A...420...79B} and in the classical BALQSO LBQS 2212-1759 \citep{2006A&A...446..439C}.

It should be noted that recent observations of radio-loud BALQSOs, thought to be seen face-on on the basis of radio emission variability, have found little or no X-ray absorption, so suggesting a different scenario for polar outflows \citep{2008ApJ...676L..97W}. 

Thanks to the current availability of both optical and X-ray large catalogs as the Sloan Digital Sky Survey \citep[SDSS,][]{2000AJ....120.1579Y} and the Second {XMM-{\textit{Newton}}} Serendipitous Source Catalog \citep[2XMM,][]{2XMM} we have been able to make a step forward in joining the two above-mentioned strategies used in BALQSOs X-ray studies. Here we present X-ray analysis of a large (54 sources) sample of BALQSOs drew from the cross-correlation of the above mentioned two catalogs. 

A cosmology with $H_0=70\:$km~s${-1}$~Mpc$^{-1}$, $\Omega_{\Lambda}=0.7$ and $\Omega_{\rm{M}}=0.3$ is used throughout the paper.

\section{The sample}\label{sec:sample}

BALQSOs are divided in two main classes: High ionization BALQSOs (HiBALs) show blueshifted, deep absorption troughs corresponding to resonant transitions of highly ionized elements and are usually classified by means of their C~IV$\lambda 1549$ BAL profile, while Low ionization BALQSOs (LoBALs) show, in addition to the high ionization ones, absorption troughs due to less ionized elements and are habitually identified by BALs in Mg~II$\lambda 2798$ line profile. While HiBALs are found in $\sim 10-15\%$ of optically selected QSOs, LoBALs are more unusual and comprise about 1\% of optically selected QSOs. 

BALQSOs are historically identified by means of their balnicity index \citep[BI,][]{1991ApJ...373...23W}, that is a C~IV equivalent width measure modified in order to exclude, from low-resolution UV spectra, intervening absorbers and narrow intrinsic absorption lines. Specifically, BI is computed considering any absorption trough spanning $\geq 2000$ km s$^{-1}$ in width, absorbing at least 10\% of the local continuum, and blueshifted by $\geq 3000$ km s$^{-1}$ with respect to the corresponding emission lines. 
While this widely used criterion assures that all QSOs with BI$\,>0$ are really BALQSOs, it may exclude intrinsic absorption troughs if starting very near the emission lines and/or high velocity, but somewhat narrow ones. For these reasons, a less restrictive criterion to identify BALQSOs has been introduced in the last years, the absorption index \citep[AI,][]{2002ApJS..141..267H,2006ApJS..165....1T}. This index is computed in the same way as BI, but relaxing the 3000 km s$^{-1}$ blueshift criterion and considering all absorption troughs with a blueshift $> 0$~km s$^{-1}$ with respect to the corresponding emission lines, and with a width of at least 1000 km s$^{-1}$. 

It was demonstrated by \citet{2008MNRAS.386.1426K} that while the use of BI excludes some BALQSOs, the inattentive use of AI leads to an overestimate of the true fraction of BALQSOs. For 'true' BALQSOs the AI is found to have values $\gtrsim 1000\:$km~s$^{-1}$, independently from BI values. In any case, the BALQSO candidate UV spectra should always be checked by eye to confirm the presence of significant BALs. It is important to stress that every BAL profile is unique, that there exists a large range of blueshift velocities and trough widths, and that the BAL phenomenon is far from being homogeneous \citep[e.g.][]{2003AJ....126.2594R,2002ApJS..141..267H}. For example, X-ray observations of optically weak outflows (i.e. mini-BALQSOs) have often revealed very strong X-ray outflows or the presence of massive ionized absorbers, both of which often variable \citep[e.g.][]{2003ApJ...595...85C, 2007AJ....133.1849C, 2004ApJ...603..425G}.\\

We cross-correlated the Sloan Digital Sky Survey Data Release 5 Quasar Catalog \citep{2007AJ....134..102S} with the Second {XMM-{\textit{Newton}}} Serendipitous Source Catalogue \citep{2XMM} using a radius of 5 arcseconds\footnote{This corresponds to $3\sigma$ mean positional uncertainty for 2XMM sources.}. 
A total of 1067 quasars of the SDSS DR5 are included in the 2XMM catalog, i.e., they have a $0.2-12$~keV detection likelihood $> 6$. 
Given the good astrometric accuracy of 2XMM Catalog and the sky density of SDSS QSOs, the identifications are robust and we do not expect to have random coincidence of positions between the two catalogs \citep[see Section~9.5 of][]{2XMM}.
About 635 SDSS DR5 QSOs have XMM-\textit{Newton} coverage but are not detected in X-rays.

The SDSS spectroscopic window covers the wavelength range of $3800-9200\,\AA $ and so allows the search for HiBALs in the redshift range $1.7 \leq z\leq 4.38$ and LoBALs in the range $0.5 \leq z \leq 2.15$.
Among the 1067 QSO detected, 945 lie at $0.5 \leq z \leq 4.38$ and therefore are searchable for BALQSOs. Among these, 203 QSOs are in the redshift range allowing for the search of both Lo- and HiBALs, while 151 (591) lie in the redshift range where only HiBALs (LoBALs) can be found.
We choose to use both BI and AI criteria to select our BALQSOs candidates. 
33 out of 945 sources are classified as BALQSOs according to \citet{2006ApJS..165....1T} using both $\rm{BI} > 0$ and $\rm{AI} > 0$ criteria. To these we added 21 sources classified as BALQSOs according to \citet{2008ApJ...680..169S} which use only the classical $\rm{BI} > 0$ criterium, so obtaining a sample of 54 BALQSOs. 
Here we should stress that the selection criteria used by \citet{2008ApJ...680..169S} are somewhat different, and less restrictive, from the ones used by \citet{2006ApJS..165....1T}: when analyzing the UV spectrum in order to measure the absorption strength (i.e. AI and/or BI), the former use a smoothing of 15 pixels, while the latter use a smoothing of 3 pixels.

It is interesting to note that the number of X-ray detected sources corresponds to about one third of BALQSOs belonging to both \citet{2006ApJS..165....1T} and \citet{2008ApJ...680..169S} catalogs and with XMM-\textit{Newton} coverage, the remaining two thirds being non-detected.

Of the 54 BALQSOs of the sample, only four sources are target of pointed observations, namely 0911+0550 \citep[PI: J.~Hjorth, see][]{2005MNRAS.364..195P}, 1141-0143 (PI: P.~Hall), 1525+5136 \citep[PI: N.~Brandt, see][]{2005AJ....130.2522S} and 1543+5359 \citep[PI: S.~Mathur, see][]{2003AJ....126.1159G}.
Among the selected 54 BALQSOs there are only 22 sources for which we have X-ray statistics good enough to obtain moderate-quality spectral measures via direct spectral fitting, i.e., $\gtrsim 100$ counts detected by EPIC-pn instrument in the $0.2-10$~keV energy band. For the remaining 32 sources, we perform an X-ray hardness ratio analysis.

\begin{figure}
\resizebox{\hsize}{!}{\includegraphics{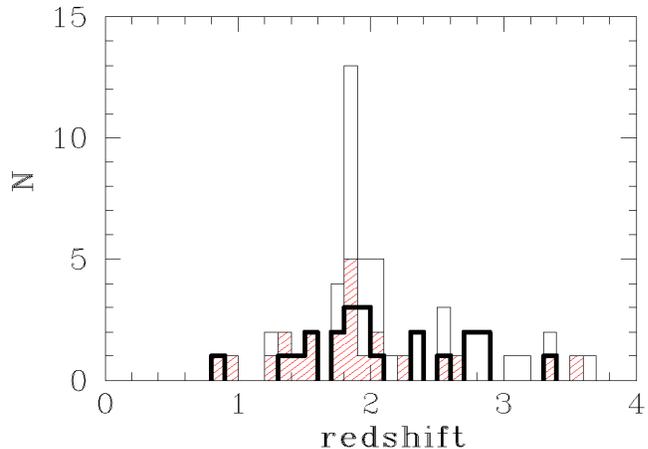}}
\caption{Redshift distribution of the 54 sources analyzed in this work. Shaded histogram refer to the 21 BALQSOs taken from \citet{2008ApJ...680..169S}, while empty histogram refer to the 33 BALQSOs taken from \citet{2006ApJS..165....1T}. Thick histogram indicates the 22 sources on which we perform spectral analysis.}
\label{fig:redshift}
\end{figure}

In Table \ref{table:1} we present the sample analyzed in this work. The horizontal line separates the BALQSOs taken from \citet{2006ApJS..165....1T}, upper table, from the BALQSOs taken from \citet{2008ApJ...680..169S}, lower table. The first column lists the nomenclature for the sources used throughout the paper. In the next columns the redshift, the SDSS and 2XMM names, and the offset in arcseconds between the positions given in the two catalogs are reported for each source. Then we report notes on individual sources (i.e. a ``L'' for lensed BALQSOs, a ``R'' for radio-loud BALQSOs) and a letter which specifies the kind of X-ray analysis performed (i.e. a ``H'' for hardness ratio analysis, a ``S'' for spectral analysis).
About a half of the sources have an offset less than 1$''$; only 8 sources have an offset larger than 2$''$ (the largest offset being $\sim 4.1''$).
The sample analyzed in this work spans a redshift $0.8\lesssim z\lesssim 3.7$ (see Fig. \ref{fig:redshift}) with a median $z_{\rm{med}} \sim 1.9$ and a mean $\langle z \rangle \sim 2.1$.

\section{Observations and data reduction}

We retrieved the observation data files (ODF) for each observation from the public {XMM-{\textit{Newton}}} science archive (XSA).
We reduced the data with the science analysis software (SAS) v7.0.0 and the latest calibration files, generated from June to July 2008.

Light curves at $E > 10$~keV were inspected, and high background flaring events were filtered out using the \texttt{xmmlight\_clean} script\footnote{Available at \url{http://www.sr.bham.ac.uk/xmm2/scripts.html}} to recursively remove all time bins which deviate more than $3\sigma$ from the mean count-rate. 

Bad pixels and hot columns were removed, and only good quality events (FLAG=0 for EPIC-pn, \#XMMEA\_EM for  EPIC-MOS) were retained. Given the low fluxes involved pile-up does not affect the data, and we therefore used single and double events for EPIC-pn (PATTERN$\leq$4), up to quadruple events for EPIC-MOS (PATTERN$\leq$12).

Table \ref{table:2} lists the observation log for the 22 sources with spectral analysis available: the first column reports the name of the source, next come the observation ID, the nominal duration of the observation, the net exposure time after the flare filtering was applied, the local background-subtracted count-rate and the corresponding errors reported in parentheses for each instrument used in the spectral analysis.  

\section{Data analysis}

Individual source spectra were generated for each instrument using circular extraction regions, with radii determined with the \texttt{eregionanalyse} package in order to maximize the signal-to-noise ratios (S/N). 
Typical extraction radii are 15--30$''$ for the source spectra, 30--60$''$ for the background spectra. 
Background was extracted from source-free regions as near as possible to the source, and its area was normalized to the source's one via the \texttt{backscale} task. 
If the source was found to lie near the edge of a CCD or near a hot column, the exposure was discarded. 

We generated ancillary response file (ARF) and redistribution matrix file (RMF) at the position of each source with the \texttt{arfgen} and \texttt{rmfgen} tasks. 
Given the current calibration uncertainties, we extracted events with energies in the range $0.2-10$~keV band in the case of EPIC-pn, $0.3-8$~keV band for the two EPIC-MOS.

We grouped the spectra at 15 counts/bin and propagated errors using Churazov weight \citep{1996ApJ...471..673C} to properly account for the low number of counts per bin that approximate a Gaussian distribution in order to apply $\chi^2$ statistics during minimization.
We checked our results with respect to other grouping (e.g. 20 counts/bin) and other statistics \citep[e.g. C-statistics,][]{1979ApJ...228..939C} and found similar results, the main difference being in the errors estimation.
We choose to group our spectra with 15 counts/bin for the following reasons:
\begin{itemize}
\item photoelectric cutoff moves at lower energies in the observer frame with increasing redshift, so we need the maximum accuracy in determining the spectral shape at lowest energies in order to measure neutral intrinsic absorption. This is not achieved with 20 counts/bin because of the low number of lowest-energy spectral bins. For the faintest sources this grouping results in  badly constrained parameters.\\
\item when using C-stat to model ungrouped spectra we obtained the same results, with errors even smaller than the 15 counts/bin case, but strongly background-modeling dependent. Being the EPIC background a relevant component of the signal collected, it can not be neglected, so we choose to subtract it and we finally used $\chi^2$ statistics.
\end{itemize}

\subsection{X-ray spectral Analysis}\label{subsec:analysis}

We used Heasoft package v6.4, Xspec v12.4.0 \citep{1996ASPC..101...17A}.
All models include appropriate Galactic neutral absorption \citep{1990ARA&A..28..215D}. Photoionization cross-sections are given by \citet{1992ApJ...400..699B}, while abundances are solar, given by \citet{1989GeCoA..53..197A}.
All quoted errors are at 90\% confidence level for one parameter of interest \citep[i.e. $\Delta\chi^2=2.706$,][]{1976ApJ...210..642A} except otherwise stated. 
    
We started modeling the spectra with a redshifted powerlaw continuum emission absorbed by Galactic neutral hydrogen.
For each source we fitted simultaneously the spectra of all instruments available, keeping only the powerlaw normalizations free in order to account for the cross-calibration uncertainties between EPIC-pn and EPIC-MOS, that are around 5-7\% (M.~Kirsch 2007\footnote{XMM-SOC-CAL-TN-0018,\url{http://xmm2.esac.esa.int/docs/documents/CAL-TN-0018.pdf}}). All model parameters are kept linked in the case of the two MOS.
    
We then added an absorption component at the redshift of the source and registered the $\chi^2$ improvement. 
We found the addition of a neutral redshifted absorption component to be statistically significant at $> 90\%$ confidence level in only 13/22 sources; among these, only for 6 sources (0043+0052, 0243+0000, 0911+0550, 1425+3757, 1525+5136 and 1543+5359) the fit improvement is significant at $> 99\%$ confidence level.
For the other 9/22 sources we found only upper limits on the amount of intrinsic neutral absorption.
Overall, 7/22 sources show $N_{\rm{H}}<4\times 10^{21}\:$cm$^{-2}$ at the 90\% confidence level.
In order to further investigate this issue, we varied for each source $\Gamma$ and $N_H$ around their best-fit values and generated contours of iso-$\Delta\chi^2=2.30, 4.61$ and 9.21 corresponding to 68, 90 and 99\% confidence level for two parameters of interest.
    
Spectral analysis results are reported in Table \ref{table:3}.
For each source we list Galactic neutral hydrogen column density $N_{\rm{H, Gal}}$, photon index $\Gamma$, neutral hydrogen column density at the redshift of the source $N_{\rm{H}}$, the powerlaw normalization N$_{\rm{1keV}}$, and the best-fit value in terms of $\chi^2$/degrees of freedom, (d.o.f.).
Next we computed the soft band ($0.5-2\:$keV) and hard band ($2-10\:$keV) observed fluxes, corrected only for Galactic absorption, and $2-10\:$keV rest-frame luminosities corrected for both Galactic and intrinsic absorption.
Spectra for each source are reported in Appendix \ref{appendix:1} together with confidence contours for $\Gamma$ and $N_{\rm{H}}$, where the vertical line marks the Galactic neutral hydrogen column density value.
    
\subsection{X-ray hardness ratio analysis}\label{sec:hr}

We performed a hardness ratio analysis on the 32 BALQSOs with the worst statistics, using EPIC-pn data only. We extracted source counts from circular regions centered at the source positions, and background counts from source-free circular regions as near as possible to the source. We generated for source and background spectra the corresponding ARFs and RMFs to account for the spatial dependence of effective area and hence of counts number. 
We computed the off-axis corrected hardness ratio between $0.2-2$ and $2-8$~keV using count-rates derived from Xspec. Errors on hardness ratios are propagated from count-rate errors using the numerical formula of \citet{lyons}. The number of counts in the $0.2-10$~keV band is approximately $30\div60$.

\begin{figure}
\centering
\includegraphics[width=8 cm]{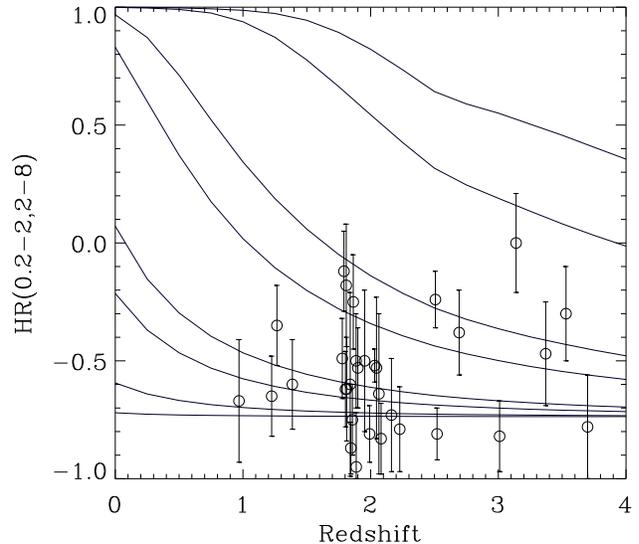}
\caption{Observed hardness ratio and 1$\sigma$ associate errors for the 32 BALQSOs with the worst statistics. The continuous tracks represent simulated hardness ratio for different redshift and column density values which are, from bottom to top: $N_{\rm{H}}=(0.01,0.1, 0.5, 1.0, 5.0, 10, 50, 100)\times 10^{22}$~cm$^{-2}$.\label{fig:hr}}
\end{figure}

We simulated the observed hardness ratio for a set of fake sources modeled with a redshifted powerlaw with $\Gamma=2$, absorbed by a varying amount of neutral hydrogen at the source redshift, and spanning a redshift range of $0\leq z \leq 4$. We then compared the observed hardness ratios with the expected ones and deduced the appropriate amount of absorbing column density required to reproduce the data (see Fig.~\ref{fig:hr}). Hardness ratio measurements, their 1$\sigma$ errors, and the corresponding absorbing column densities are reported in Table~\ref{table:3b}. 

\subsection{Optical/X-ray photometry}
    
We measured Galactic absorption corrected, source rest-frame flux densities $f_{\rm{2keV}}$ and $f_{\rm{2500}}$ from the {XMM-{\textit{Newton}}} and SDSS spectra, respectively. 
In order to measure $f_{\rm{2keV}}$ we de-redshifted the $0.5-2\:$keV fluxes using the best-fit photon indices listed in Table~\ref{table:3}. In the case of the 32 sources without spectral analysis, we estimated the $0.5-2\:$keV rest-frame fluxes from the observed count-rate using \texttt{webpimms}\footnote{\url{heasarc.gsfc.nasa.gov/Tools/w3pimms.html }} with the appropriate effective area, redshift, and a fixed photon index $\Gamma=2$.
For the $f_{\rm{2500}}$ measurements we used the five SDSS magnitudes and followed the method outlined in Section 2.2 of \citet{2003AJ....125..433V}.
    
We then computed $\alpha_{\rm{ox}}=0.384\log f_{\rm{2keV}}/\log f_{\rm{2500}}$, the spectral slope of a hypothetical powerlaw connecting the $2500\,\AA$ and $2\:$keV rest-frame flux densities \citep{1979ApJ...234L...9T}. Several studies handled the $\alpha_{\rm{ox}}$ distribution in QSOs and found a strong correlation with rest-frame $2500\AA$ luminosity density $l_{\rm{2500}}$. We compared our observed $\alpha_{\rm{ox}}$ with the most updated\footnote{During the acceptance stage of this paper, \citet{2008arXiv0808.2603G} published a new $l_{\rm{2500}}-\alpha_{\rm{ox}}$ correlation, so actually updating the \citet{2007ApJ...665.1004J} one. Albeit the new correlation of \citet{2008arXiv0808.2603G} is somewhat steeper than the one used in this work, it does not affect significantly the present conclusions.}  best-fit correlation $l_{\rm{2500}}-\alpha_{\rm{ox}}$ for SDSS radio-quiet QSOs \citep[Equation 3 of][]{2007ApJ...665.1004J}. 

Results and measurements are reported in Table~\ref{table:4} together with other optical properties. Specifically, we report also the AI and $v_{\rm{out}}$ (i.e. the absorption index and the maximum velocity outflow) measured for each source: when available, we used the AI/$v_{\rm{out}}$ values given in \citet{2006ApJS..165....1T}, otherwise we estimated them from SDSS spectra and we flag these values with an asterisk. The BAL subclassification is also given in the table: we use Hi/Lo to mark HiBALs/LoBALs, and H to mark those BALQSOs for which the Mg~II spectral region is redshifted outside the SDSS spectroscopic window, so preventing us from checking the presence of low-ionization absorption troughs.

\section{Results}

\begin{figure}
\centering
\resizebox{\hsize}{!}{\includegraphics{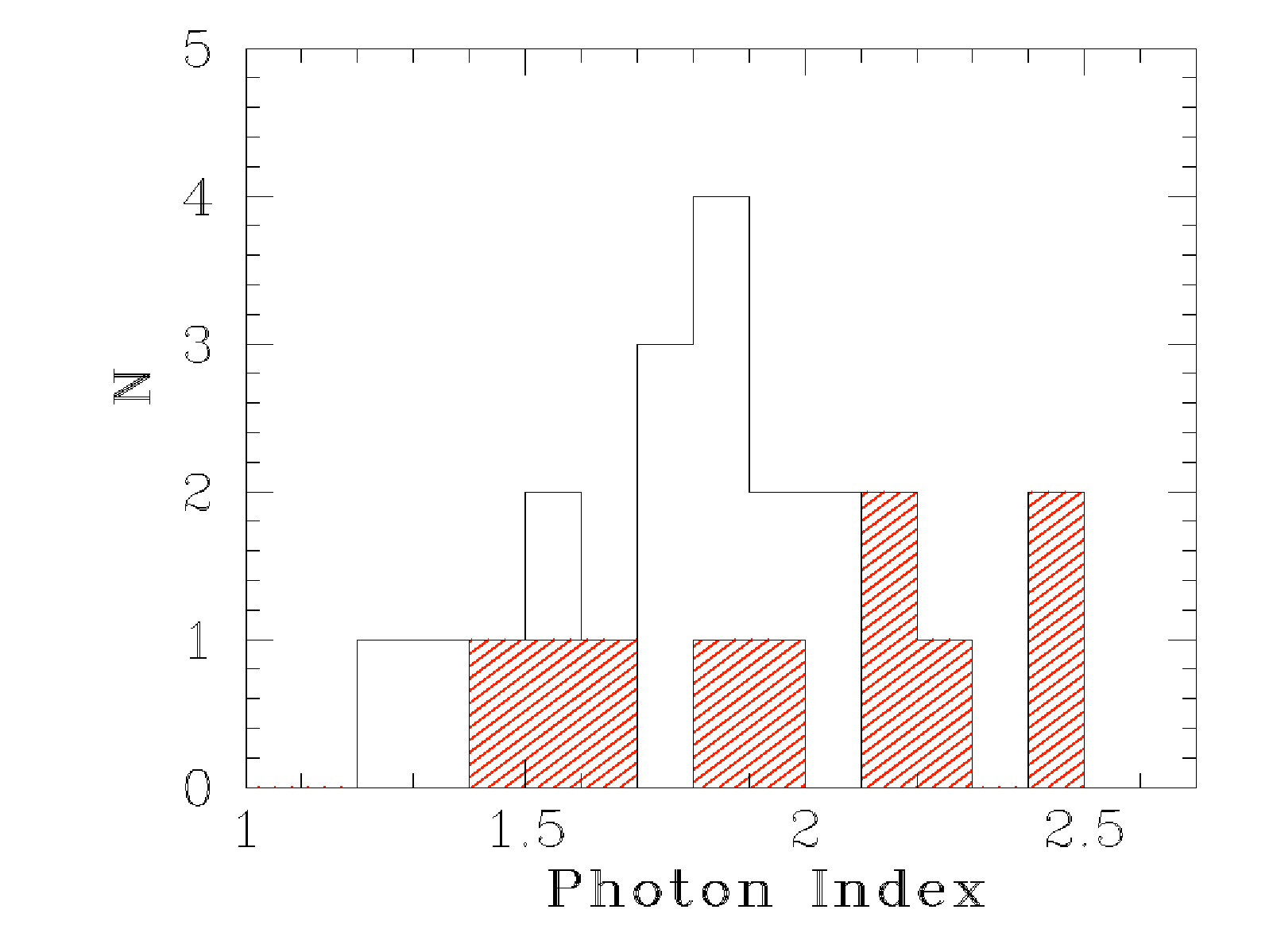}}
\resizebox{\hsize}{!}{\includegraphics{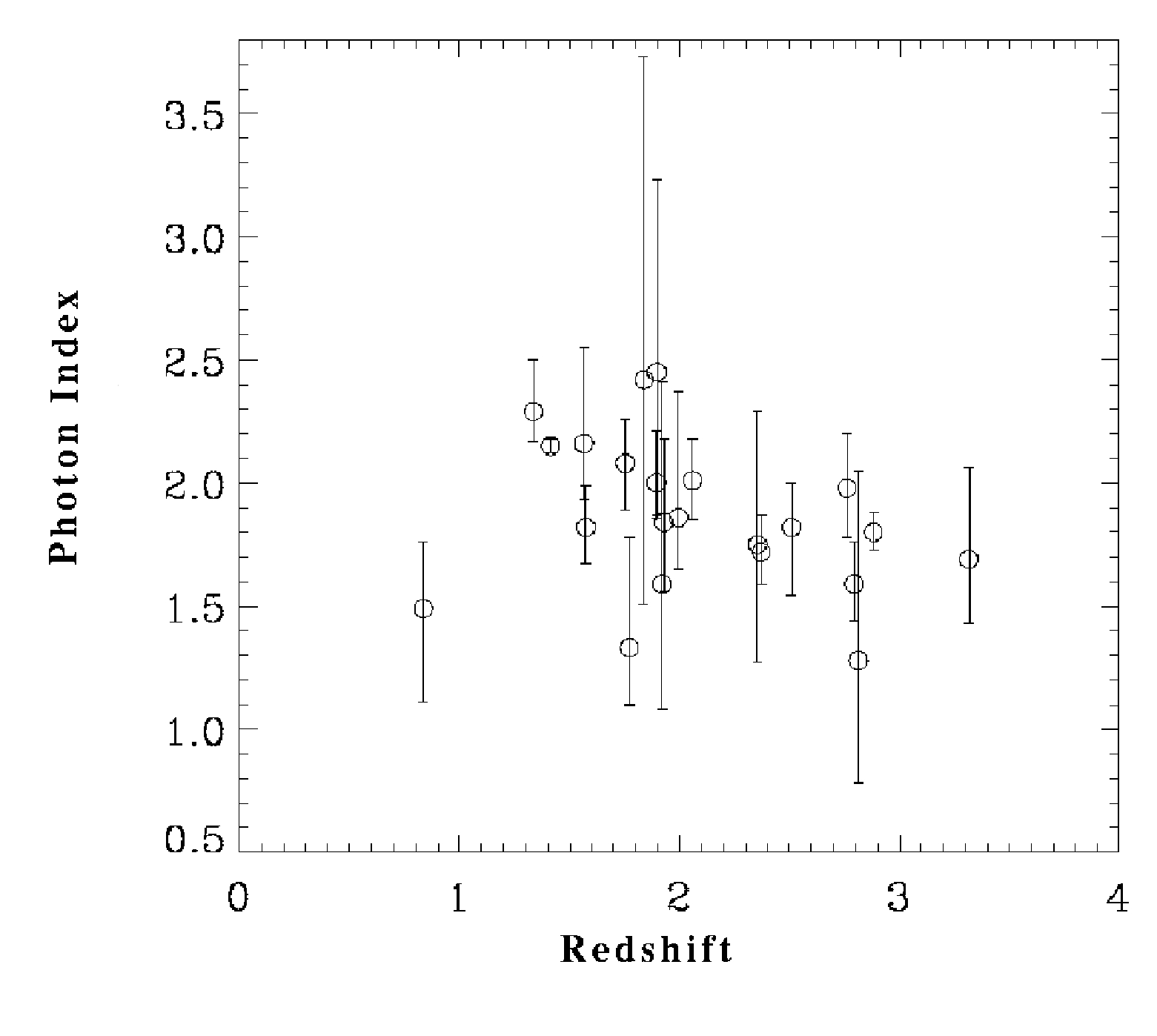}}
\caption{\label{fig:phindex}Top panel: photon index distribution for the 22 BALQSOs with spectral analysis available; shaded histogram refers to the BALQSOs taken from \citet{2008ApJ...680..169S}. Bottom panel: measured photon indices versus redshift.} 
\end{figure}

The best-fit photon index distribution is shown in Fig.~\ref{fig:phindex}, top panel, for the 22 BALQSOs on which we performed spectral analysis.
The sample mean photon index is $\langle \Gamma\rangle=1.87$ with a dispersion $\sigma=0.21$, thus compatible with typical radio-quiet type 1 AGN \citep[e.g.][]{2005A&A...432...15P}; there are no differences among the distributions of the two subsamples taken from \citet{2008ApJ...680..169S} and from \citet{2006ApJS..165....1T} and introduced in Section~\ref{sec:sample}.
A couple of sources (namely 1007+5343 and 1011+5541) present rather flat photon indices $\Gamma \sim 1.3$ that could indicate the presence of more complex (e.g. ionized/partially covering) absorbers.
No significant evolution with redshift is seen (see Fig.~\ref{fig:phindex}, bottom panel).

\begin{figure}
\centering
\resizebox{\hsize}{!}{\includegraphics{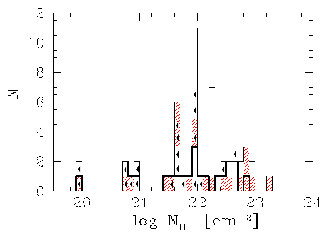}}
\resizebox{\hsize}{!}{\includegraphics{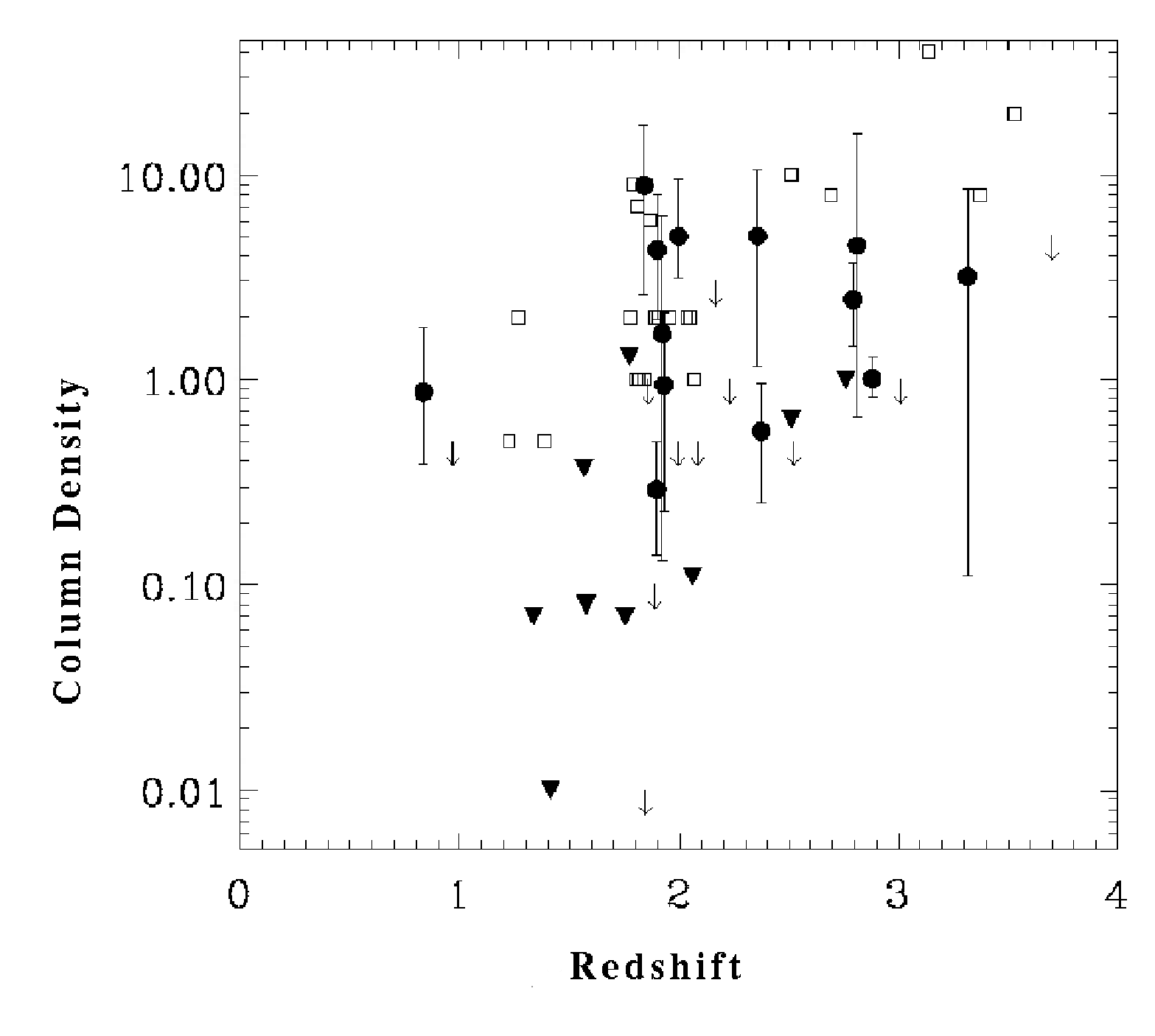}}
\caption{\label{fig:nh}Top panel: neutral column density distribution for the 22 BALQSOs with spectral analysis (thick histograms) and for the 32 BALQSOs with hardness ratio analysis (thin histograms) available. Shaded histogram refers to the BALQSOs taken from \citet{2008ApJ...680..169S}, left-pointing arrows represent upper limits. Bottom panel: column densities in units of $10^{22}$~cm$^{-2}$ versus redshift: filled symbols refer to the 22 BALQSOs with spectral analysis available, where downward-pointing triangles are the upper limits at the 90\% confidence level. Empty symbols refer to the 32 BALQSOs with hardness ratio available, where arrows are upper limits at the 68\% confidence level.}    
\end{figure}

The intrinsic neutral absorber column density distribution is shown in Fig.~\ref{fig:nh}, top panel, for the BALQSOs analyzed through both spectral and hardness ratio analysis, where for the latter class we fixed the photon index at $\Gamma=2$ (see Section~\ref{sec:hr}).
While we expected a $\langle N_{\rm{H}} \rangle \sim 10^{23}\:$cm$^{-2}$ (see Section~\ref{sect:intro}), we found $ N_{\rm{H}} \leq 10^{22}\:$cm$^{-2}$ for 28/54 sources. Again, there are no differences among the distributions of the two BALQSO subsamples  introduced in Section~\ref{sec:sample}.
We checked $N_{\rm{H}}$ dependence on redshift (see Fig.~\ref{fig:nh}, bottom panel) and found no significant correlation, except a shallow tendency to measure higher $N_{\rm{H}}$ values with increasing redshift consistent with being due to the shift of the photoelectric cutoff outside of the observed energy band \citep[see e.g. discussions in][]{2006A&A...451..457T, 2006A&A...459..693A}.
We wondered whether the low measured column densities are related to the low number of spectral bins often used in spectral analysis, but since we found no correlation among the number of counts and the measured column densities, we conclude that this is not the case.
We stress here that intrinsic column densities might be much higher if either the underlying continuum is more complex than a simple power law (e.g. if there is a significant contribution from soft-excess and/or a scattered component) or if the absorber is more complex than simple neutral one (e.g. if it is ionized and/or partially covering the continuum emission source). 

\begin{figure}
\centering
\resizebox{\hsize}{!}{\includegraphics{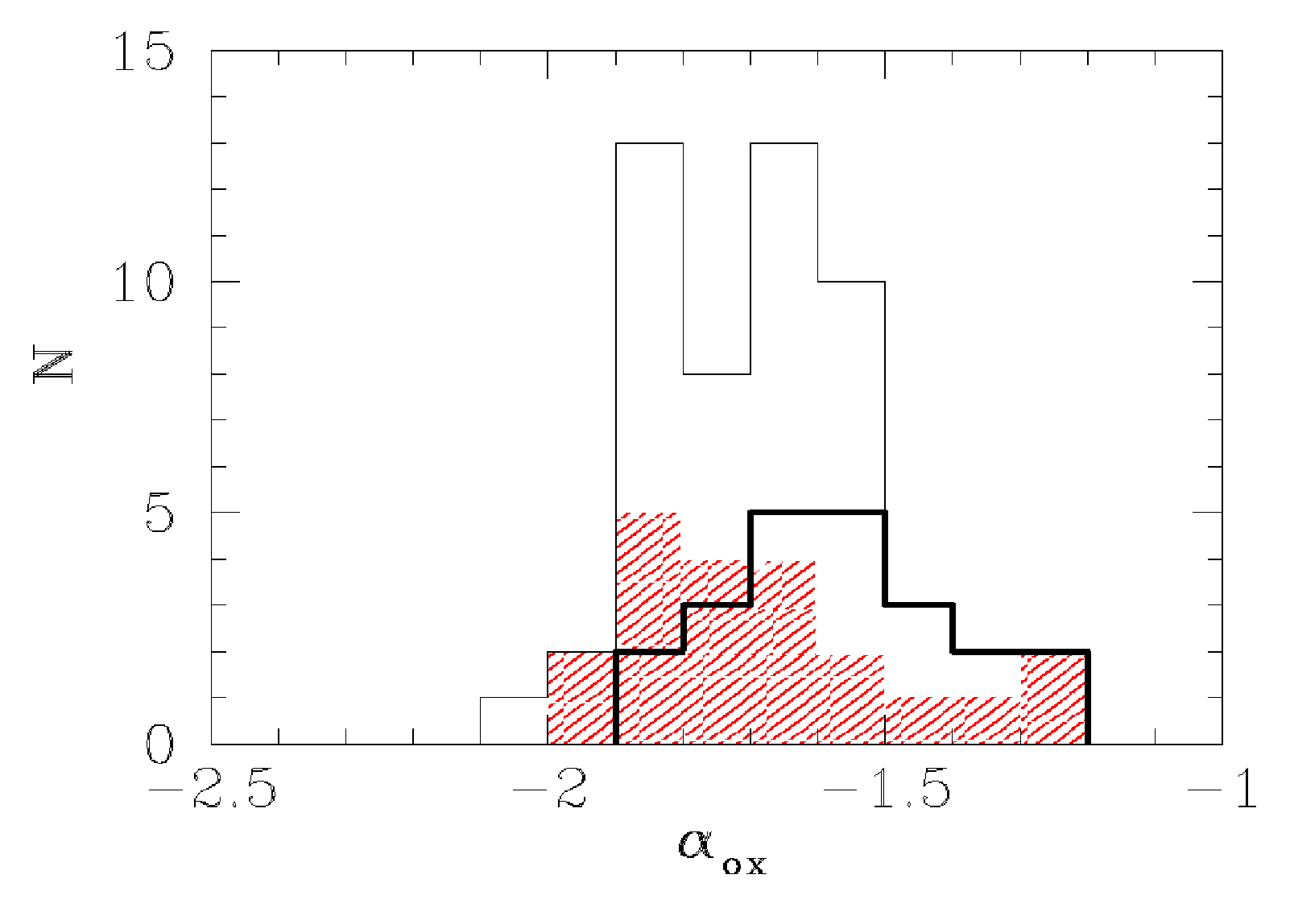}}
\resizebox{\hsize}{!}{\includegraphics{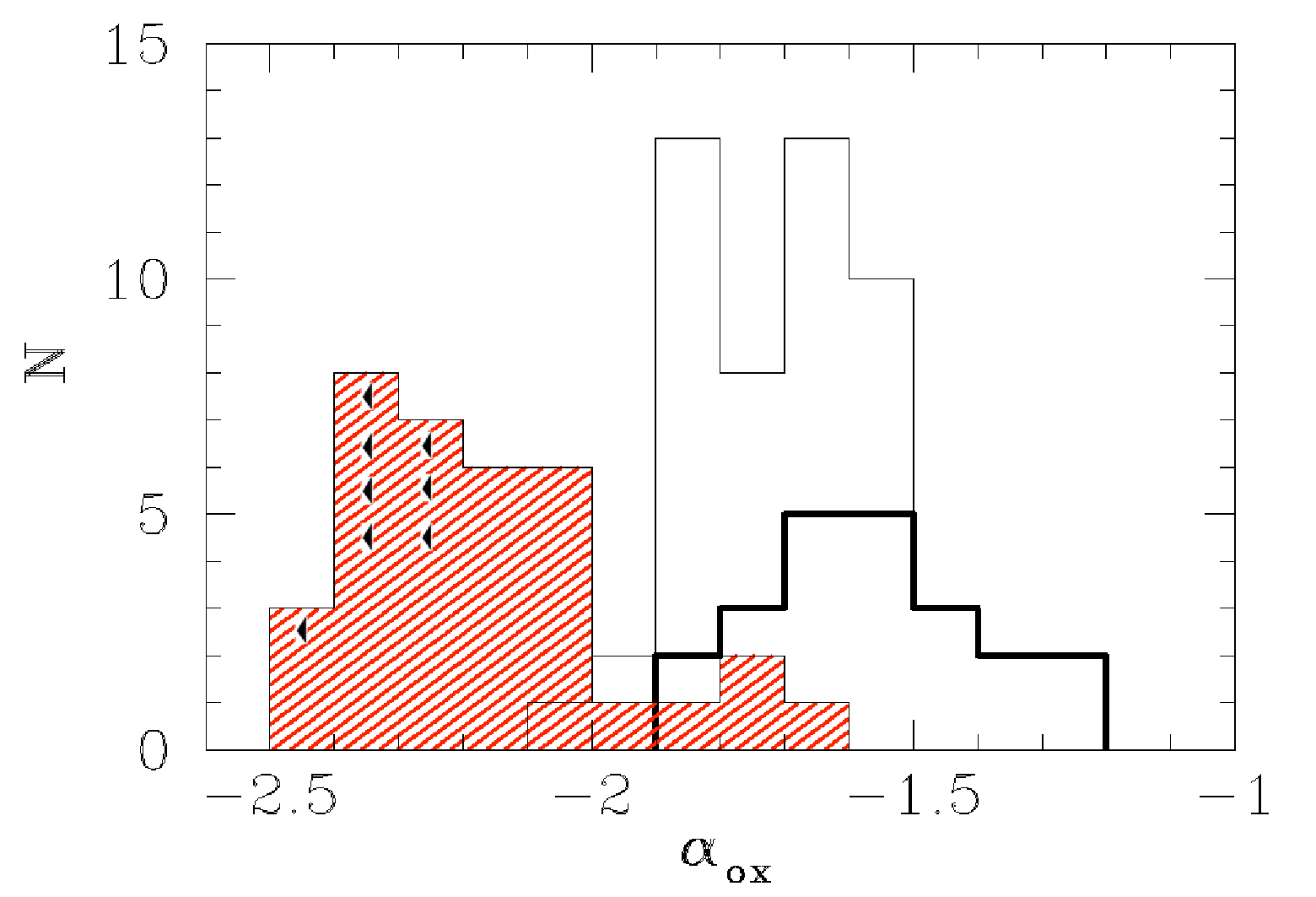}}
\caption{Top panel: $\alpha_{\rm{ox}}$ distribution for our 54 SDSS BALQSOs, where shaded histogram refers to the 21 BALQSOs taken from \citet{2008ApJ...680..169S} and thick histogram refers to the 22 sources on which we performed spectral analysis. Bottom panel: Comparison between $\alpha_{\rm{ox}}$ measured for our 54 SDSS BALQSOs and for the 35 LBQS BALQSOs (shaded histograms) studied by \citet{2006ApJ...644..709G}.}
\label{fig:aox}
\end{figure}

\begin{figure}
\centering
\resizebox{\hsize}{!}{\includegraphics{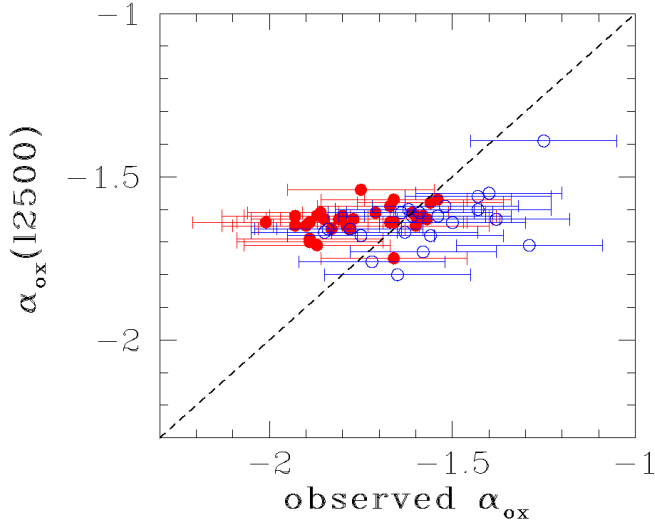}}
\caption{Measured $\alpha_{\rm{ox}}$ versus  $\alpha_{\rm{ox}}(l_{\rm{2500}})$, where open (blue) circles are the 22 sources on which we performed X-ray spectral analysis, filled (red) circles are the 32 sources on which we performed X-ray hardness ratio analysis.}
\label{fig:aox2}
\end{figure}

The observed $\alpha_{\rm{ox}}$ distribution is shown in Fig.~\ref{fig:aox}, top panel. 
Once more, there are no evident differences among the distribution of the two BALQSOs subsamples. The observed distribution is typical of radio-quiet, unabsorbed broad-line (i.e. type 1) AGN and atypical of BALQSOs which usually present an observed $\alpha_{\rm{ox}} \sim -(2\div 2.5)$ \citep[e.g.][]{1996ApJ...462..637G, 2006ApJ...644..709G}, see Fig.~\ref{fig:aox}, bottom panel. 
Our measured $\alpha_{\rm{ox}}$ are plotted in Fig. \ref{fig:aox2}, where the dashed line represent the $\alpha_{\rm{ox}}(l_{\rm{2500}})$ value expected on the basis of rest-frame $2500\,\AA$ luminosity density according to Eq.~3 of \citet{2007ApJ...665.1004J}. 
The subsample with X-ray spectra seems to show a slightly flatter $\alpha_{\rm{ox}}$ (open circles, $\langle \alpha_{\rm{ox}} \rangle \sim -1.57$) than the hardness ratio analyzed one (filled circles, $\langle \alpha_{\rm{ox}} \rangle \sim -1.75$), but according to a Kolmogorov-Smirnov statistic test (KS) this difference is not statistically significant (less than 2$\sigma$). 
None but one of our BALQSOs is found to lie in the typical region of ``soft X-ray weak'' QSOs, that is $\alpha_{\rm{ox}}<-2$, but nearly all of them have an observed $\alpha_{\rm{ox}}$ consistent, within the errors, with the $\alpha_{\rm{ox}}(l_{\rm{2500}})$.

\section{Discussion}

Two remarkable results of our analysis are the observed $\alpha_{\rm{ox}}$ distribution, that is atypical of BALQSOs, and the low intrinsic $N_{\rm{H}}$ value measured for more than a quarter of the sample (15/54 sources), $N_{\rm{H}} < 5\times 10^{21}\:$cm$^{-2}$. We discuss below in turn each of these issues, which both point independently to an unabsorbed scenario for a large fraction of our BALQSO sample. The most meaningful comparisons can be made with respect to the largest X-ray studied BALQSO sample known to date, that is the \citet{2006ApJ...644..709G} {\textit{Chandra}} study of 35 Large Bright Quasar Survey \citep[LBQS,][]{1995AJ....109.1498H} BALQSOs.

While \citet{2006ApJ...644..709G} find a mean $\Delta\alpha_{\rm{ox}}\equiv \alpha_{\rm{ox}} - \alpha_{\rm{ox}}(l_{\rm{2500}}) \sim -0.53$, we obtain $\langle \Delta \alpha_{\rm{ox}} \rangle \sim -0.04$, i.e. a difference of a factor of 3 in UV/X-ray luminosity ratio.
Assuming the same amount of mean absorption for the two samples, the smaller $|\Delta\alpha_{\rm{ox}}|$ measured in our sample with respect to the LBQS BALQSO could be due to a lower optical luminosity and/or to a higher X-ray luminosity of our sample with respect to LBQS BALQSO.
A KS statistic test on the rest-frame $2500\,\AA$ luminosity densities yields a probability $< 3\times 10^{-3} $ that the two samples are drawn for chance from the same UV parent population, with the LBQS BALQSOs being, on average, more UV luminous than SDSS BALQSOs.
While our BALQSOs are selected from the same parent sample used by \citet{2007ApJ...665.1004J} in their study, that is mainly based on spectroscopically confirmed, optically selected SDSS QSOs, the LBQS sample used by \citet{2006ApJ...644..709G} comprises extremely bright and luminous, color selected QSOs.
It may be that the LBQS BALQSO have a higher UV/X-ray luminosity ratio than the typical QSOs and so the $\alpha_{\rm{ox}}-l_{\rm{2500}}$ correlation may not hold.
In any case, it should be stressed that this correlation has a high intrinsic dispersion $\left(\Delta(\alpha_{\rm{ox}})\sim 0.2\right)$ and has been constructed excluding BALQSOs, therefore these considerations remain speculative.
On the other hand, in this work we are dealing with X-ray selected (i.e., on the basis of the 2XMM catalog) BALQSOs, and so we are biased toward the X-ray brightest ones. We may be presenting a particular, unexpected X-ray loud class of BALQSOs instrinsically more X-ray luminous than those studied so far. Among the 54 BALQSOs studied in this work, we find a significant number of sources with $\alpha_{\rm{ox}}$ less negative than \textit{absorption-corrected} \citet{2006ApJ...644..709G} values, and this result makes the above possibility a plausibly one. 

Alternatively, the main driver for the differences observed in the $\alpha_{\rm{ox}}$ distribution could be the amount of X-ray absorption.
Indeed, the most striking difference in results between LBQS BALQSOs and SDSS BALQSOs is the amount of intrinsic absorption.
\citet{2006ApJ...644..709G} estimate, from hardness ratio analysis, an X-ray absorbing column density of $\langle N_{\rm{H}} \rangle \sim 10^{23}\:$cm$^{-2}$ for all their LBQS BALQSOs, while we measure from spectral fitting and hardness ratio analysis a much lower absorption for a large fraction of our SDSS BALQSOs, and no absorption at all (in excess of the Galactic one) for more than a third of the spectroscopically analyzed sample (see Table~\ref{table:3} and Fig.~\ref{fig:graphics:a}).
A possible explanation may lie in the small bandpass of {\textit{Chandra}} (0.5-8 keV) that makes difficult for previous works to measure accurately the low-energy photoelectric cutoff with increasing redshift (the mean redshift of the two samples is comparable, $z \sim 2$) and so may bring to an overestimation of intrinsic column density \citep[see e.g. discussions and simulations in][]{2006A&A...451..457T}. On the other hand, the high effective area of {XMM-{\textit{Newton}}} EPIC-pn at lowest energies (down to 0.2~keV) make our column densities measures robust even at the highest redshift probed by our sample. 
We may be missing the ``classical'' absorbed, soft-X-ray weak BALQSOs and probing the bright tail of the known distribution. A comparison of the fraction of X-ray non-detected BALQSOs (i.e. BALQSOs covered by XMM-\textit{Newton} observations but not included in the 2XMM catalog) with the fraction of X-ray non-detected non-BALQSOs (see Section~\ref{sec:sample}), together with the smooth $\alpha_{\rm{ox}}$ distribution -- albeit with different $\langle\alpha_{\rm{ox}}\rangle$ -- among BALQSOs analyzed through spectral fitting and hardness ratio, let us
argue that this possibility may be true. We defer the study of the X-ray properties of the X-ray non-detected BALQSOs to a forthcoming paper. 
However, the significant fraction of very low intrinsic column densities and very flat $\alpha_{\rm{ox}}$ observed in our sample strongly assess the existence of some unabsorbed, X-ray bright BALQSOs.

We then explored the intriguing possibility that the observed X-ray properties are related to some physical properties of outflows. 
It is interesting to note that a high UV/X-ray luminosity ratio is requested theoretically to launch a radiatively driven UV wind up to very high velocities in order to prevent overionization (and so the uneffective acceleration) of the outflowing gas.
The ``BAL-nicity'' (quantified by either BI or AI) of \citet{2006ApJ...644..709G} sample is much stronger than ours, indicating somewhat faster, more powerful and more massive outflows observed in their LBQS BALQSO sample (BI is a coarse measure of the convolution of minimum and mean velocity of the outflowing UV absorbing gas, and of its amount along the line of sight).
We may be probing the low-velocity tail of BALs, that tend to become overionized by the strong X-ray continuum and so cause a weaker UV outflow to be observed.

\begin{figure}[h!]
\centering
\resizebox{\hsize}{!}{\includegraphics{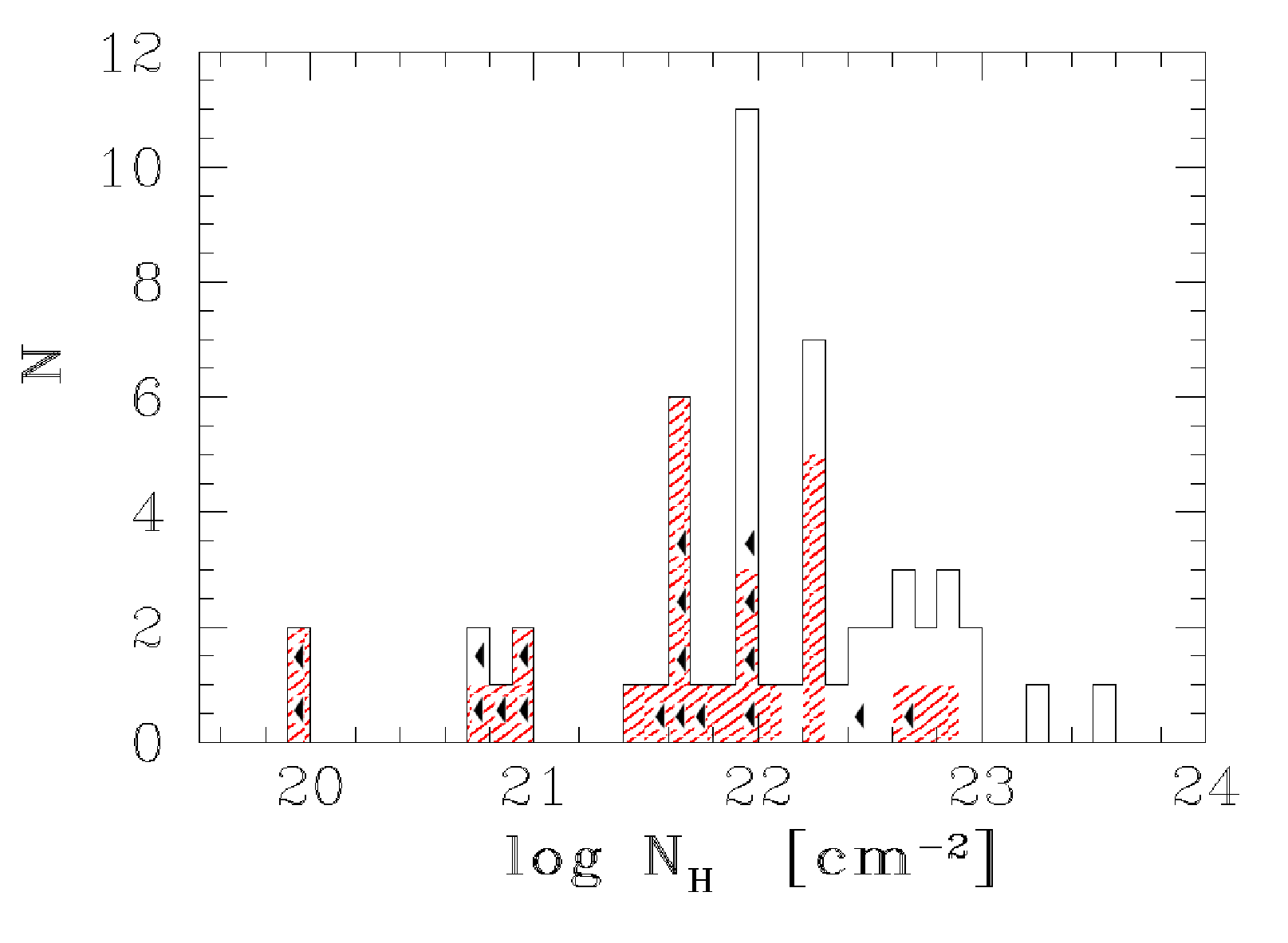}}
\resizebox{\hsize}{!}{\includegraphics{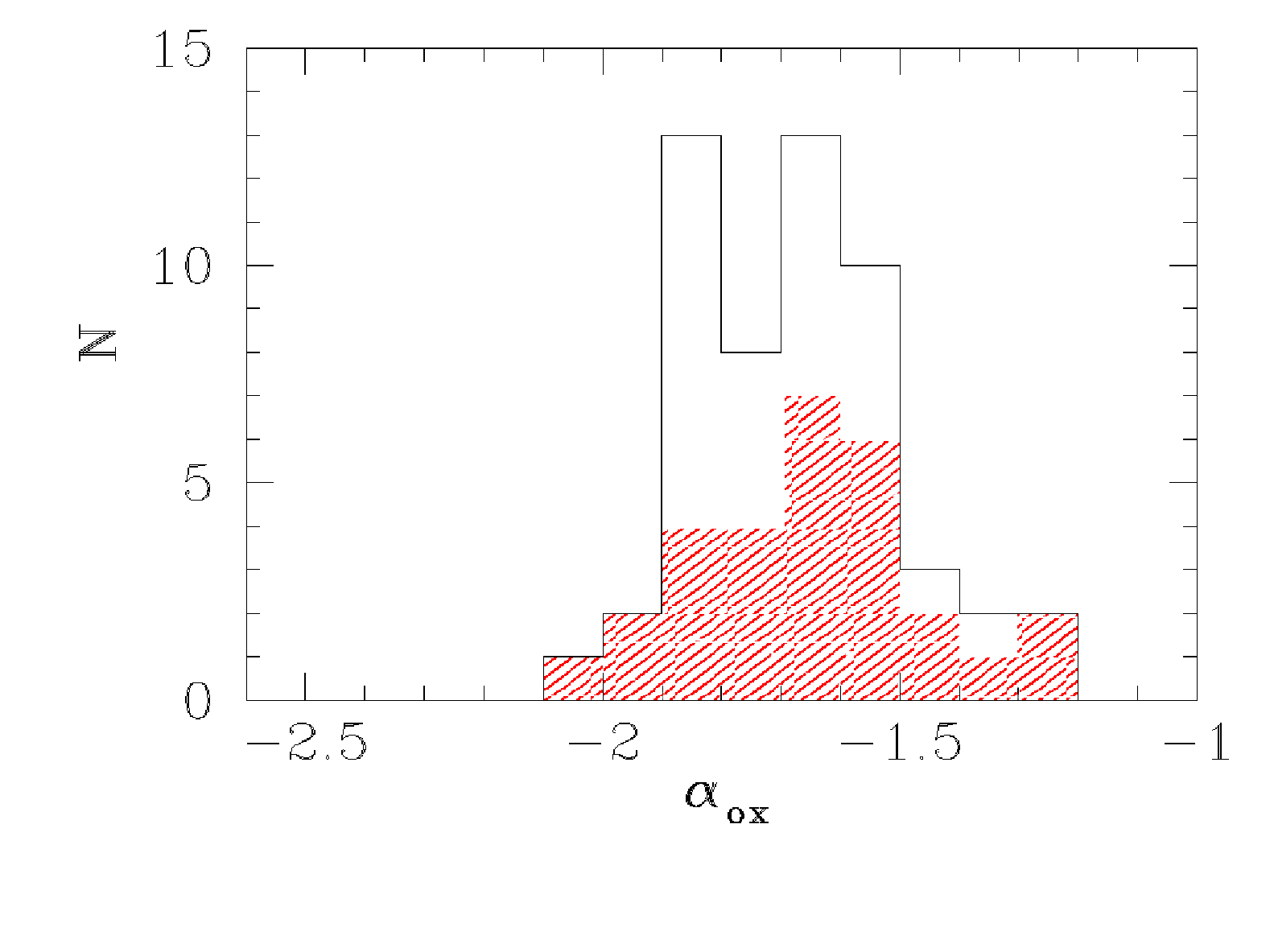}}
\caption{Distribution of neutral absorbing column densities (top panel) and $\alpha_{\rm{ox}}$ (bottom panel), where the shaded histograms refer to the 30 BALQSOs with $\langle\rm{AI}\rangle\sim 500$~km s$^{-1}$ and the empty histograms refer to the 24 BALQSOs with $\langle\rm{AI}\rangle\sim 2000$~km s$^{-1}$. Arrows indicate upper limits.}
\label{fig:ai-nh}
\end{figure}

\begin{figure}[h!]
\centering
\resizebox{\hsize}{!}{\includegraphics{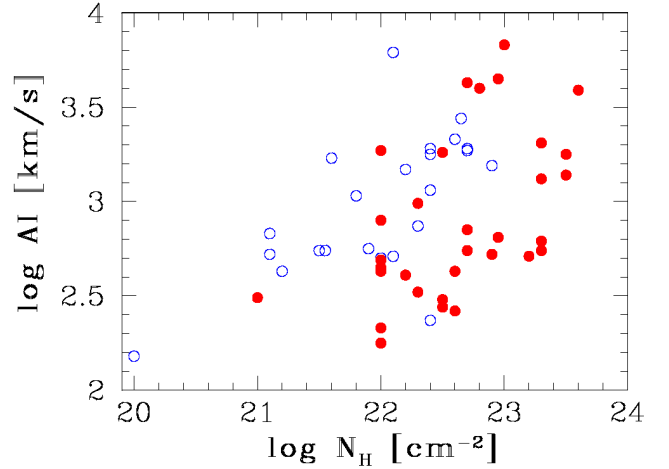}}
\caption{Absorption Index versus measured X-ray absorbing column density for the BALQSOs with spectral analysis available (open circles) and with hardness ratio analysis available (filled circles).}
\label{fig:ai2}
\end{figure}

We divided our BALQSO sample in low- and high-AI subsamples according to \citet{2008MNRAS.386.1426K}.
We find 30/54 sources belonging to the low-AI group ($\langle \rm{AI}\rangle \sim 500~$km s$^{-1}$) and 24/54 sources belonging to high-AI group ($\langle \rm{AI}\rangle \sim 2000$~km s$^{-1})$. 
In Fig. \ref{fig:ai-nh} we re-plot the $N_{\rm{H}}$ and $\alpha_{\rm{ox}}$ distributions where this time the shaded histogram refers to the low-AI BALQSOs. It seems that the low-AI subsample tends to occupy the low-$N_{\rm{H}}$ tail of the column density distribution, for which only one high-AI BALQSO presents $N_{\rm{H}}<5\times 10^{21}$~cm$^{-2}$, the other 14 unabsorbed BALQSOs being low-AI. No differences are evident among the $\alpha_{\rm{ox}}$ distributions of low- and high-AI BALQSOs.
We used the Spearman's $\rho$ bivariate non-parametric statistic test, implemented into ASURV rev 1.2 \citep{1992ASPC...25..245L}, in order to search for correlation among AI and $N_{\rm{H}}$ accounting for upper limits in the column density measurements. We find a probability of 0.0026 that AI and $N_{\rm{H}}$ are not correlated. Although this value can not be considered as a strong proof of the existence of a correlation between the two quantitities, it strenghtens the conclusion that the high-AI BALQSO subsample appears to be more X-ray absorbed than the low-AI one.
We plot in Fig.~\ref{fig:ai2} measured AI versus measured $N_{\rm{H}}$ for all the 54 sources of our sample.
Previous works \citep[i.e.][]{2008MNRAS.386.1426K} have shown that there seems to exist a subsample of low-AI BALQSOs that are usually included in BALQSOs samples even if their UV properties are statistically different from those of classical BALQSOs. Our work confirms the different behavior of these sources also from an X-ray perspective. 

Even considering only the high-AI BALQSO subsample, we find rather low X-ray absorbing column densities, and high $\alpha_{\rm{ox}}$, with respect to the expectations based on known literature results and theoretical disk wind models.

\begin{figure}[h!]
\centering
\resizebox{\hsize}{!}{\includegraphics{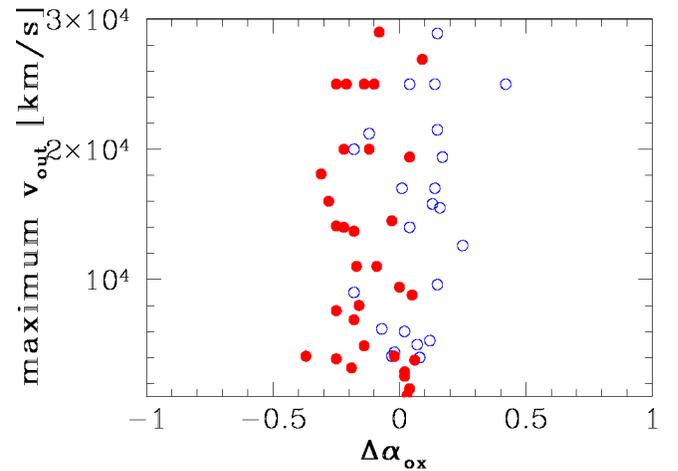}}
\caption{$\Delta\alpha_{\rm{ox}}$ versus maximum velocity UV outflow for the BALQSOs with spectral analysis available (open circles) and with hardness ratio analysis available (filled circles).} 
\label{fig:vout-dox}
\end{figure}

We checked whether a correlation between $\Delta\alpha_{\rm{ox}}$ and maximum velocity outflow does exist for our SDSS BALQSOs, as found by \citet{2006ApJ...644..709G} for their LBQS BALQSOs. This correlation is usually interpreted as a signature of radiatively driven accretion disk wind, because of the association of the highest velocity UV outflows with the more negative $\Delta\alpha_{\rm{ox}}$  (i.e. with the greater amount of X-ray absorption). This is easily interpreted as the presence of an effectively shielding column density of X-ray absorbing gas that prevents the UV wind from continuum overionization and makes possible to radiatively accelerate it to the highest velocities observed \citep[see e.g.][]{2007ASPC..373..305G}.
Fig.~\ref{fig:vout-dox} shows that such correlation does not hold for our BALQSOs sample.
We checked whether the lack of correlation between $\Delta\alpha_{\rm{ox}}$ and maximum velocity outflow is driven by the inclusion of the low-AI BALQSOs in the sample, but this is not the case.
The same lack of correlation between $\Delta\alpha_{\rm{ox}}$ and maximum velocity outflow also holds for the NALQSOs sample of \citet{2008ApJ...677..863M} and is interpreted by the authors as a different location above the accretion disk of the gas responsible for the UV NALs (higher latitudes above the accretion disk) and BALs (lower latitudes), with UV and X-ray absorbers being different in both cases.  
Some of our BALQSOs are indeed quite precisely classified as mini-BALQSOs, that manifest intermediate UV properties between NALQSOs and BALQSOs. Mini-BALQSOs are usually thought to be related to the same physical phenomena of BALQSOs -- the difference being in the line of sight along the UV outflow -- and indeed we found similar characteristics for their X-ray absorbers and for the other BALQSOs of our sample. 

Some of the BALQSOs selected from \citet{2008ApJ...680..169S}, namely 1001+5553, 1021+1315 and 1447+4032, may be not genuine BALQSOs nor mini-BALQSOs, and they could have been included in the BALQSO sample because of the large smoothing applied by the authors to the UV spectra in their analysis. Even if some of these sources could be NALQSOs -- or even if their UV spectrum could ``only'' bring the signature of intervening absorbers -- our conclusions remain unchanged, that is, X-ray selected BALQSOs are not soft X-ray weak and they appear less X-ray absorbed than purely optically selected BALQSOs.

The low absorption found in our BALQSO sample recalls the results found by \citet{2008ApJ...676L..97W} for a little sample of radio-loud BALQSOs. While in this case the authors argue for either a wind launching mechanism different than for ``classical'' BALQSOs (e.g. magnetic pressure driven winds linked to the presence of radio jets) or for the presence of absorbers more complex than totally covering, neutral ones, we can not reach any conclusion about these hypoteses because of the low number of formally radio-loud\footnote{According to the \citet{1989AJ.....98.1195K} definition, that is $f_{\rm{5GHz}}/f_{4400\AA}>10$, where 5 GHz and 4400\AA\ are in the source rest-frame.} BALQSOs in our sample (that is, 4/54, namely 0923+5127, 1141-0143, 1146+4723 and 1324+0320) and for the low statistics that prevents us from constraining some physical parameters such as the ionization state and/or the covering fraction of the absorbing gas.
Furthermore, if the presence of a significant soft-excess holds for our sample as found by previous studies in a high fraction of radio-quiet QSOs \citep[e.g.][]{2005A&A...432...15P}, we could be severely underestimating the actual intrinsic column density. Unfortunately, the low statistics again is preventing us from checking this hypothesis.
The above issues suggest that our measured column densities can be safely considered as lower limits, but we note that given the redshift distribution of our sources, for most of our BALQSOs the soft-excess would be redshifted outside the observed band. However, we stress that previous BALQSOs studies found significantly higher column densities than ours, without account for the soft-excess issue.

It is also worth noting that some AGN -- both BALQSOs \citep[see e.g.][]{2007A&A...474..431S, 2008A&A...483..137B} and non-BALQSOs \citep[see e.g.][]{2007ApJ...668L.111G, 2007AJ....133.1988G} -- with multiple and deep X-ray observations have revealed a transient behaviour in terms of soft X-ray weakness, related to an X-ray variability stronger than the UV one. While this could be the reason for the observed X-ray properties of some sources of our sample (if they have been catched in a ``high'' X-ray state), the large number of BALQSOs studied in this work rule out this possibility as the only explanation for the observed UV/X-ray properties of the full sample.  

Overall, our results can be reconciled with theoretical accretion disk wind models if we are looking at the SDSS/2XMM BALQSOs along lines of sight which are not nearly equatorial as they are thought to be in the case of ``classical'' BALQSOs, but rather at smaller angles with respect to the accretion disk rotation axis, so missing both the bulk of X-ray shielding gas and the highest velocities, highest column densities UV outflows. If this scenario holds, we can conclude that the outflow properties of QSOs smoothly connect each other in every kind of QSO, the observed differences being driven by purely geometrical effects.
This scenario can be tested by studying the X-ray properties of the remaining $\sim 100$ X-ray non-detected BALQSOs taken from the SDSS DR5 QSOs catalog.

\section{Conclusions}

We analyzed X-ray properties of 54 BALQSOs drawn from the cross-correlation of SDSS DR5 and 2XMM catalogs by performing spectral analysis on 22 sources, hardness ratio analysis on the remaining ones.

Our results in terms of $\alpha_{\rm{ox}}$ and the amount of intrinsic X-ray absorption are different with respect to the literature results, that are mainly based on purely optically selected BALQSOs.
We found none but one of our X-ray selected BALQSOs being soft X-ray weak, but the most striking result of our spectral analysis is the low intrinsic X-ray absorbing column density value measured for more than a half of the sample, $N_{\rm{H}} \leq  10^{22}\:$cm$^{-2}$.

We observationally confirm the statistical result presented by \citet{2008MNRAS.386.1426K} which found a bimodal distribution in AI for BALQSOs; we found that the low intrinsic neutral X-ray absorption measured for a large fraction of the sample is linked to the low AI measured from their UV spectra. We should treat with caution such kind of sources when included in BALQSO samples until we do not clarify their role in the quasar outflow scenarios. Overall, we found a lower than expected $\langle N_{\rm{H}}\rangle$ also for the high-AI BALQSO subsample.

The low measured intrinsic absorption may be reconciled with theoretical results by allowing either the intrinsic continuum to comprise a significant soft-excess or a scattered component, or the X-ray absorber to be ionized, or only partially covering the continuum source. Unfortunately, the rather low statistics does not allow us to assess the significance of the presence of an intrinsic soft-excess or of a partially covering absorber. Furthermore, the redshift of our sources prevents us from checking the presence of the most usual X-ray spectral features due to warm absorbers, i.e. the O~VII and O~VIII absorption edges. Investigating the presence of more highly-ionized gas (e.g. absorption features due to Fe~XXV, Fe~XXVI), absorbing the X-ray continuum at the highest energies probed by our observations, would need much deeper observations of such distant sources with the current X-ray telescopes. 
With these caveats in mind, we can conclude that our measured intrinsic column densities might be sometimes lower limits to the presence of intrinsic absorption for this BALQSO sample.

We may be probing either a tail of the known BALQSO population, probably related to a line of sight with larger angle with respect to the accretion disk plane, or a different outflow launching mechanism that does not need a large amount of X-ray absorbing, shielding gas to accelerate the wind, i.e. a different than radiatively driven quasar wind or a scenario which includes magnetic forces as shielding and/or accelerating factors. The relative number of X-ray non-detected BALQSOs compared to the number of X-ray non-detected SDSS DR5 QSOs, together with the smooth connection of X-ray properties between bright (i.e. spectrally analyzed) and faint (i.e. hardness ratio analyzed) X-ray detected BALQSOs strongly suggests the former hypothesis. If this were true, we can expect more X-ray absorption in the $\sim 100$ BALQSOs with only X-ray upper limits.
These intriguing results deserve to be investigated more accurately with the increase of the number of BALQSOs well-studied in both the UV and X-ray band in order to test the accretion disk wind models and the quasar outflow scenarios.

\begin{acknowledgements}
We thank G.G.C.~Palumbo for thoughtful discussions and fundamental support, G.~Chartas and M.~Eracleous for key and pleasant discussions, Y.~Shen for providing her electronical SDSS DR5 BALQSO catalog, S.~Mateos and M.~Watson for useful clarifications about the 2XMM catalog, G.~Risaliti for interesting discussions occurred during this work, G.~Zamorani for a careful reading of an early manuscript and for punctual objections and subsequent discussions, M.~Dadina for continuous collaboration and discussions, M.~Mignoli for help in dealing with UV measurements, R.~Della~Ceca, P.~Severgnini, A.~Caccianiga and S.C.~Gallagher for very useful discussions. We also would like to thank the anonymous referee for useful comments that significantly improved the paper.

The authors thank the Italian Space Agency for financial support (contracts ASI--INAF I/023/05/0 and ASI I/088/06/0).

Funding for the SDSS and SDSS-II has been provided by the Alfred P. Sloan Foundation, the Participating Institutions, the National Science Foundation, the U.S. Department of Energy, the National Aeronautics and Space Administration, the Japanese Monbukagakusho, the Max Planck Society, and the Higher Education Funding Council for England. The SDSS Web Site is \url{http://www.sdss.org/}.

\end{acknowledgements}

\bibliographystyle{aa}
\bibliography{bibbal}

\Online

\begin{appendix}
\section{Tables}\label{appendix:tables}
\begin{table*}[!h]
\caption{SDSS/2XMM BALQSO sample}
\label{table:1}
\centering          
\begin{tabular}{c c c c c c c }
\hline\hline
Name & $z$ & SDSS name  & 2XMM name & Offset  & Notes & Analysis\\
 & & & & [$''$] &      & \\
(1) & (2) & (3) & (4) & (5) & (6) & (7) \\
\hline
0042-0912 & 1.778 & SDSS J004206.18-091255.7 & 2XMM J004206.1-091255 & 0.78 &   & H \\
0231-0739 & 2.508 & SDSS J023148.80-073906.3 & 2XMM J023148.8-073905 & 0.36 &   & S\\
0243+0000 & 1.995 & SDSS J024304.68+000005.4 & 2XMM J024304.6+000005 & 0.54 &    & S\\
0734+3203 & 2.082 & SDSS J073405.24+320315.2 & 2XMM J073405.2+320315 & 0.78 & &  H\\
0855+3757 & 1.929 & SDSS J085551.24+375752.2 & 2XMM J085551.1+375752 & 0.60 &  &  S\\
0911+0550 & 2.793 & SDSS J091127.61+055054.0 & 2XMM J091127.5+055054 & 0.36 & L &  S \\
0921+5131 & 1.845 & SDSS J092142.57+513149.4 & 2XMM J092142.5+513148 & 1.32 & &  H\\
0922+5121 & 1.753 & SDSS J092238.43+512121.2 & 2XMM J092238.3+512120 & 0.72 &    & S\\
0923+5127 & 2.163 & SDSS J092345.19+512710.0 & 2XMM J092345.1+512711 & 1.14 & R & H \\
1001+0140 & 2.059 & SDSS J100116.78+014053.5 & 2XMM J100116.7+014053 & 1.68 &    & S\\
1001+0224 & 2.032 & SDSS J100145.15+022456.9 & 2XMM J100145.2+022456 & 1.08 & & H \\
1007+5343 & 1.772 & SDSS J100728.69+534326.7 & 2XMM J100728.8+534327 & 1.86 & L &  S\\
1011+5541 & 2.813 & SDSS J101144.33+554103.1 & 2XMM J101144.4+554103 & 1.02 &    & S\\
1019+0825 & 3.010 & SDSS J101954.54+082515.0 & 2XMM J101954.6+082515 & 2.16 &  & H \\
1106+5222 & 1.840 & SDSS J110637.16+522233.4 & 2XMM J110637.0+522233 & 1.08 &  & H \\
1123+0524 & 3.699 & SDSS J112300.25+052451.0 & 2XMM J112300.3+052448 & 2.46 &  & H \\
1135+4913 & 1.992 & SDSS J113537.67+491323.2 & 2XMM J113537.6+491322 & 0.24 &  & H \\
1141-0143 & 1.266 & SDSS J114111.61-014306.6 & 2XMM J114111.5-014305 & 1.26 & R  & H \\
1223+1034 & 2.761 & SDSS J122307.52+103448.2 & 2XMM J122307.4+103448 & 0.12 &    & S\\
1227+0126 & 1.953 & SDSS J122708.29+012638.4 & 2XMM J122708.2+012638 & 0.30 & &  H \\
1236+6158 & 2.520 & SDSS J123637.45+615814.4 & 2XMM J123637.7+615813 & 1.26 & &  H \\
1245-0021 & 2.354 & SDSS J124520.72-002128.2 & 2XMM J124520.6-002127 & 0.48 &    & S\\
1324+0320 & 3.371 & SDSS J132401.53+032020.6 & 2XMM J132401.4+032019 & 1.08 & R &  H \\
1328+5818 & 3.139 & SDSS J132827.07+581836.9 & 2XMM J132827.3+581839 & 3.78 & &  H \\
1340-0019 & 1.857 & SDSS J134059.24-001944.9 & 2XMM J134059.1-001945 & 1.32 & & H \\
1435+4841 & 1.886 & SDSS J143513.89+484149.3 & 2XMM J143513.9+484149 & 0.78 &  & H \\
1446+0255 & 1.864 & SDSS J144625.48+025548.6 & 2XMM J144625.6+025549 & 1.80 &  & H \\
1508+5652 & 1.803 & SDSS J150858.15+565226.5 & 2XMM J150858.2+565227 & 1.14 &  & H \\
1517+0016 & 1.887 & SDSS J151729.70+001652.6 & 2XMM J151729.6+001651 & 1.08 &  & H \\
1525+5136 & 2.882 & SDSS J152553.89+513649.1 & 2XMM J152553.8+513649 & 0.60 &   & S\\
1533+3243 & 1.900 & SDSS J153322.80+324351.4 & 2XMM J153322.8+324351 & 0.60 &  & H \\
1543+5359 & 2.370 & SDSS J154359.44+535903.2 & 2XMM J154359.4+535902 & 0.36 &   & S\\
2039-0102 & 2.065 & SDSS J203941.04-010201.6 & 2XMM J203941.2-010202 & 3.72 &  & H \\
& & & & & & \\
\hline
& & & & & & \\
0043+0046 & 1.574 & SDSS J004338.10+004615.9 & 2XMM J004338.0+004616 & 0.48 &   & S\\
0043+0052 & 0.834 & SDSS J004341.24+005253.3 & 2XMM J004341.3+005253 & 1.02 &    & S\\
0109+1328 & 1.225 & SDSS J010941.97+132843.8 & 2XMM J010941.8+132844 & 1.62 & &  H \\
0242-0000 & 2.507 & SDSS J024230.65-000029.6 & 2XMM J024230.6-000030 & 1.32 &  & H\\
0302+0006 & 3.315 & SDSS J030222.08+000631.0 & 2XMM J030222.0+000630 & 0.90 &  & S\\
0729+3700 & 0.969 & SDSS J072945.33+370031.9 & 2XMM J072945.3+370032 & 0.54 &  & H\\
0919+3030 & 1.387 & SDSS J091914.23+303019.0 & 2XMM J091914.2+303018 & 0.42 &  & H\\
0939+3556 & 2.046 & SDSS J093918.07+355615.0 & 2XMM J093918.1+355612 & 2.64 &  & H \\
0943+4811 & 1.809 & SDSS J094309.56+481140.5 & 2XMM J094309.8+481142 & 3.42 &  & H \\
1001+5553 & 1.413 & SDSS J100120.84+555349.5 & 2XMM J100120.7+555351 & 1.32 & L  & S\\
1021+1315 & 1.565 & SDSS J102117.74+131545.9 & 2XMM J102117.7+131546 & 0.78 &   & S\\
1052+4414 & 1.791 & SDSS J105201.35+441419.8 & 2XMM J105201.3+441417 & 1.62 &  & H \\
1124+3851 & 3.530 & SDSS J112432.14+385104.3 & 2XMM J112432.0+385104 & 1.02 &  & H \\
1146+4723 & 1.895 & SDSS J114636.88+472313.3 & 2XMM J114636.9+472313 & 0.42 & R  & S\\
1205+4431 & 1.921 & SDSS J120522.18+443140.4 & 2XMM J120522.1+443141 & 0.90 &   & S\\
1335+5147 & 1.838 & SDSS J133553.61+514744.1 & 2XMM J133553.7+514744 & 1.20 &   & S\\
1336+5146 & 2.228 & SDSS J133639.40+514605.2 & 2XMM J133639.1+514608 & 4.14 &  & H \\
1425+3739 & 2.693 & SDSS J142555.22+373900.7 & 2XMM J142555.2+373900 & 0.54 &  & H \\
1425+3757 & 1.897 & SDSS J142539.38+375736.7 & 2XMM J142539.3+375736 & 0.54 &   & S\\ 
1426+3753 & 1.812 & SDSS J142652.94+375359.9 & 2XMM J142652.8+375401 & 2.10 & &  H \\
1447+4032 & 1.335 & SDSS J144727.49+403206.3 & 2XMM J144727.4+403206 & 0.60 &   & S\\ 
\hline
\end{tabular}\\
\begin{flushleft}
\small{Notes: Col.(1): Source name used throughout the paper; Col.(2): Redshift, taken from SDSS; Col.(3): SDSS source name; Col.(4): 2XMM source name; Col.(5): Offset between the positions given in the two catalogs; Col.(6): Notes on individual objects: L $=$ lensed QSO, R $=$ Radio-loud QSO; Col.(7): Kind of X-ray analysis used in this work: S $=$ spectral analysis, H $=$ hardness ratio analysis. The horizontal line separates BALQSOs taken from  \citet{2006ApJS..165....1T} (upper table) from BALQSOs taken from \citet{2008ApJ...680..169S} (lower table).}
\end{flushleft}
\end{table*}

\begin{table*}
\caption{X--ray observations log for the 22 sources with Spectral Analysis}
\label{table:2}
\centering          
\begin{tabular}{c c c c c}  
\hline\hline
 Name & OBSID & Duration & Exposure & Count-rate \\
      &       &          & M1/M2/pn & M1/M2/pn \\
      &       &  [ks]    & [ks]     &  [ct ks$^{-1}$]\\
(1) & (2) & (3) & (4) & (5)\\
\hline
0231-0739 & 0200730401 & 44.9 & 41.7/42.4/33.7 & 1.41(0.24)/1.58(0.22)/7.33(0.68)  \\
0243+0000 & 0111200101 & 42.2 & 36.7/34.3/--   & 3.02(0.30)/2.84(0.31)/--\\
          & 0111200201 & 46.4 & 33.5/32.3/--   & 2.73(0.33)/2.55(0.35)/--\\
0855+3757 & 0302581801 & 29.1 & 27.2/27.1/17.8 & 2.09(0.29)/1.55(0.27)/7.19(0.85)  \\
0911+0550 & 0083240201 & 20.7 & 19.3/19.2/13.4 & 10.59(0.86)/12.48(0.90)/34.75(2.18) \\
0922+5121 & 0300910301 & 40.9 & 37.6/37.8/33.2 & 3.41(0.39)/3.05(0.38)/11.80(1.08)\\
1001+0140 & 0302351001 & 43.5 & 42.5/42.0/33.9 & 2.23(0.26)/2.61(0.28)/11.43(0.81)  \\
1007+5343 & 0070340201 & 34.2 & 21.7/21.6/16.5 & 1.87(0.42)/3.50(0.53)/9.01(0.95) \\
1011+5541 & 0085170101 & 33.2 & --/--/24.1     & --/--/3.65(0.51)  \\
1223+1034 & 0108860101 & 22.1 & 20.3/20.4/15.3 & 5.31(0.64)/5.72(0.62)/22.75(1.46) \\
1245-0021 & 0110980201 & 58.2 & 56.4/56.2/47.8 & 0.58(0.14)/0.82(0.16)/3.68(0.38) \\
1525+5136 & 0011830201 & 38.7 & 29.9/30.1/23.7 & 22.21(0.91)/20.71(0.89)/79.60(1.97)  \\
1543+5359 & 0060370901 & 34.1 & 24.8/23.3/21.5 & 12.41(0.81)/12.66(0.84)/45.20(2.33)  \\
 & & & & \\
\hline
 & & & & \\
0043+0046 & 0090070201 & 21.1 & 20.3/20.3/16.3 & 3.15(0.45)/5.02(0.56)/17.56(1.30)\\
0043+0052 & 0090070201 & 21.1 & --/--/16.3     & --/--/13.55(1.02)\\
0302+0006 & 0041170101 & 51.7 & 46.4/45.9/37.1 & 1.76(0.21)/1.31(0.21)/4.10(0.42)\\
1001+5553 & 0147760101 & 44.4 & 30.9/32.3/25.8 & 56.45(1.41)/54.19(1.34)/311.40(4.02) \\
1021+1315 & 0146990101 & 21.9 & 19.5/19.8/15.3 & 4.29(0.52)/5.15(0.61)/19.02(1.22) \\
1146+4723 & 0047540601 & 28.6 & 26.2/27.0/20.0 & 9.11(0.72)/9.57(0.71)/35.60(1.65) \\
1205+4431 & 0156360101 & 52.2 & --/--/18.7     & --/--/4.75(0.78) \\
1335+5147 & 0084190201 & 48.8 & --/--/35.3     & --/--/2.88(0.37) \\
1425+3757 & 0112230201 & 25.8 & 24.0/24.5/18.4 & 2.20(0.34)/1.14(0.33)/6.32(0.85) \\
1447+4032 & 0109080601 & 22.0 & 19.0/19.2/16.2 & 5.35(0.63)/4.95(0.63)/20.01(1.24)\\ 
\hline
\end{tabular}\\
\begin{flushleft}
\small{Notes: Col.(1): Source name; Col.(2): Observation ID; Col.(3): Nominal duration of the observation; Col.(4): Net exposure time for each instrument after the background flaring filtering was applied; Col.(5): Net count-rate for each instrument after the local background subtraction was applied, with corresponding errors reported in parentheses.}
\end{flushleft}
\end{table*}

\begin{table*}
\caption{Spectral analysis results}
\label{table:3}
\centering          
\begin{tabular}{ c c c c c c c c c}
\hline\hline
Name & $N_{\rm{H, Gal}}$ & $\Gamma$ & $N_{\rm{H}}$ & N$_{\rm{1keV}}$ & $\chi^2$/dof & $\log f_X(0.5-2)$ & $\log f_X(2-10)$ & $\log L(2-10)$\\
     & [$10^{20}$ cm$^{-2}$] &        & [$10^{22}$ cm$^{-2}$] &  [$10^{-4}$ ph keV$^{-1}$ cm$^{-2}$ s$^{-1}$] &           & [erg s$^{-1}$ cm$^{-2}$] & [ erg s$^{-1}$ cm$^{-2}$]     & [erg s$^{-1}$] \\
(1) & (2) & (3) & (4) & (5) & (6) & (7) & (8) & (9) \\
\hline 
0231-0739 & 3.14 & 1.82$^{+0.18}_{-0.28}$ & $<0.64$ & 0.62$^{+0.39}_{-0.15}$ & 50/31 & -13.8 & -13.7 & 45.0 \\
 & & & & & & & \\
0243+0000 & 3.56 & 1.86$^{+0.51}_{-0.21}$ & 5.02$^{+4.48}_{-1.92}$ & 1.14$^{+2.28}_{-0.46}$ (*)  & 30/26 & -13.8 (*)& -13.2 (*)& 45.0 (*)\\
 & & & & & & & \\
0855+3757 & 2.92 & 1.84$^{+0.34}_{-0.28}$ & 0.94$^{+1.17}_{-0.71}$ & 0.50$^{+0.45}_{-0.19}$ & 10/15 & -14.0 & -13.7 & 44.6 \\
 & & & & & & & \\
0911+0550 & 3.64 & 1.59$^{+0.17}_{-0.15}$ & 2.45$^{+1.23}_{-0.99}$ & 2.71$^{+1.67}_{-0.88}$ & 55/71 & -13.3 & -12.8 & 45.7 \\
 & & & & & & & \\
0922+5121 & 1.43 & 2.08$^{+0.18}_{-0.19}$ & $<0.07$ & 0.51$^{+0.10}_{-0.09}$ & 75/72 & -13.8 & -13.8 & 44.6 \\
 & & & & & & & \\
1001+0140 & 2.60 & 2.01$^{+0.17}_{-0.16}$ & $<0.11$ & 0.51$^{+0.15}_{-0.07}$ & 41/48 & -13.9 & -13.8 & 44.7 \\
 & & & & & & & \\
1007+5343 & 0.71 & 1.33$^{+0.45}_{-0.23}$ & $<1.29$ & 0.29$^{+0.41}_{-0.09}$ & 22/20 & -13.78 & -13.3 & 44.9 \\
 & & & & & & & \\
1011+5541 & 0.79 & 1.28$^{+0.77}_{-0.50}$ & 4.52$^{+11.46}_{-3.86}$ & 0.18$^{+0.81}_{-0.09}$   & 5/5  & -14.2 & -13.7 & 44.7 \\
 & & & & & & & \\
1223+1034 & 2.22 & 1.98$^{+0.22}_{-0.20}$ & $<0.99$ & 1.96$^{+1.01}_{-0.63}$ & 45/41 & -13.7 & -13.5 & 45.3 \\
 & & & & & & & \\
1245-0021 & 1.70 & 1.75$^{+0.54}_{-0.48}$ & 5.03$^{+5.52}_{-3.87}$ & 0.43$^{+0.95}_{-0.10}$ & 25/20 & -14.2 & -13.7 & 44.8 \\
 & & & & & & & \\
1525+5136 & 1.57 & 1.80$^{+0.08}_{-0.07}$ & 1.01$^{+0.28}_{-0.19}$ & 4.87$^{+0.78}_{-0.71}$ & 179/197 & -13.1 & -12.8 & 45.9\\
 & & & & & & & \\
1543+5359 & 1.25 & 1.72$^{+0.15}_{-0.13}$ & 0.56$^{+0.40}_{-0.31}$ & 0.50$^{+0.45}_{-0.19}$ & 184/147 & -13.3 & -13.0 & 45.5 \\
 & & & & & & & \\
\hline
 & & & & & & & \\
0043+0046 & 2.31 & 1.82$^{+0.17}_{-0.15}$ & $<0.08$ & 0.64$^{+0.16}_{-0.10}$ & 37/40 & -13.6 & -13.4 & 44.7 \\
 & & & & & & & \\
0043+0052 & 2.32 & 1.49$^{+0.27}_{-0.38}$ & 0.87$^{+0.92}_{-0.48}$ & 0.39$^{+0.44}_{-0.17}$ & 16/13 & -13.7 & -13.1 & 44.3 \\
 & & & & & & & \\
0302+0006 & 7.16 & 1.69$^{+0.37}_{-0.26}$ & 3.17$^{+5.33}_{-3.06}$ & 0.54$^{+0.73}_{-0.21}$ & 19/21  & -14.2 & -13.6 & 45.0 \\
 & & & & & & & \\
1001+5553 & 0.82 & 2.15$^{+0.03}_{-0.03}$ & $<0.01$ & 21.80$^{+0.60}_{-0.60}$ & 513/522& -12.1 & -12.1& 46.0 \\
 & & & & & & & \\
1021+1315 & 3.92 & 2.16$^{+0.39}_{-0.23}$ & $<0.37$ & 1.66$^{+0.83}_{-0.39}$ & 35/31 & -13.3 & -13.3 & 45.0 \\
 & & & & & & & \\
1146+4723 & 2.22 & 2.00$^{+0.21}_{-0.14}$ & 0.29$^{+0.21}_{-0.15}$ & 2.53$^{+0.80}_{-0.50}$  & 98/94 & -13.2 & -13.1 & 45.3 \\
 & & & & & & & \\
1205+4431 & 1.15 & 1.59$^{+0.82}_{-0.51}$ & 1.67$^{+4.73}_{-1.54}$ & 0.25$^{+0.91}_{-0.11}$ & 2/7 & -14.1 & -13.5 & 44.5 \\
 & & & & & & & \\
1335+5147 & 0.94 & 2.42$^{+1.31}_{-0.91}$ & 8.91$^{+8.42}_{-6.31}$ & 1.49$^{+8.51}_{-0.51}$ & 9/6 & -14.2 & -14.0 & 44.6 \\
 & & & & & & & \\
1425+3757 & 0.95 & 2.45$^{+0.78}_{-0.58}$ & 4.28$^{+3.80}_{-2.34}$ & 2.38$^{+7.79}_{-1.45}$ & 19/15  & -13.7 & -13.7 & 45.1\\
 & & & & & & & \\
1447+4032 & 1.23 & 2.29$^{+0.21}_{-0.12}$ & $<0.07$ & 1.08$^{+0.25}_{-0.12}$ & 34/36 & -13.5 & -13.6 & 44.6 \\ 
\hline
\end{tabular}\\
\begin{flushleft}
\small{Notes: Col.(1): Source name; Col.(2): Galactic neutral hydrogen column density, taken from \citet{1990ARA&A..28..215D}; Col.(3): Photon Index; Col.(4): Neutral hydrogen column density at the source redshift; Col. (5): EPIC-pn powerlaw normalization at 1 keV; Col.(6): Best-fit $\chi^2$ value over degrees of freedom; Col.(7): Logarithm of EPIC-pn $0.5-2$~keV observed flux, corrected for Galactic neutral absorption; Col.(8): Logarithm of EPIC-pn $2-10$~keV observed flux, corrected for Galactic neutral absorption; Col.(9): Logarithm of EPIC-pn $2-10$~keV rest-frame luminosity, corrected for both Galactic and intrinsic absorption. Values marked with (*) refer to EPIC-MOS instrument. }
\end{flushleft}
\end{table*}

\begin{table*}
\caption{\label{table:3b}Hardness ratio analysis results}
\centering          
\begin{tabular}{ c c c c }
\hline\hline
Name & HR & $N_{\rm{H, Gal}}$ & $N_{\rm{H}}$  \\
     &        & [$10^{20}$ cm$^{-2}$] & [$10^{22}$ cm$^{-2}$] \\
(1) & (2) & (3) & (4) \\
\hline 
0042-0912 & $-0.49\pm{0.17}$ & 2.77 & 2.0 (0.5-4.0)\\
0109+1328 & $-0.65\pm{0.17}$ & 3.26 & 0.5 (0.01-1.0)\\
0242-0000 & $-0.24\pm{0.12}$ & 2.91 & 10.0 (7.0-15.0)\\ 
0729+3700 & $-0.67\pm{0.26}$ & 5.77 & $<0.5$ \\
0734+3203 & $-0.83\pm{0.15}$ & 4.46 & $<0.5$ \\
0919+3030 & $-0.60\pm{0.19}$ & 1.64 & 0.5 (0.01-2.0) \\
0921+5131 & $-0.87\pm{0.11}$ & 1.29 & $<0.01$ \\
0923+5127 & $-0.73\pm{0.24}$ & 1.34 & $<3.0$\\
0939+3556 & $-0.53\pm{0.30}$ & 1.07 & 2.0 (0.01-6.0)\\
0943+4811 & $-0.18\pm{0.26}$ & 1.22 & 7.0 (3.0-15.0)\\
1001+0224 & $-0.52\pm{0.07}$ & 1.78 & 2.0 (1.0-3.0)\\
1019+0825 & $-0.82\pm{0.15}$ & 2.52 & $<1.0$\\
1052+4414 & $-0.12\pm{0.17}$ & 1.08 & 9.0 (5.0-15.0)\\
1106+5222 & $-0.60\pm{0.39}$ & 0.89 & 1.0 (0.01-2.0)\\
1123+0524 & $-0.78\pm{0.22}$ & 3.71 & $<5.0$ \\      
1124+3851 & $-0.30\pm{0.20}$ & 2.01 & 20.0 (7.0-30.0)\\
1135+4913 & $-0.81\pm{0.12}$ & 1.68 & $<0.5$ \\
1141-0143 & $-0.35\pm{0.17}$ & 2.16 & 2.0 (1.0-4.0)\\
1227+0126 & $-0.50\pm{0.30}$ & 1.72 & 2.0 (0.5-5.0)\\
1236+6158 & $-0.81\pm{0.11}$ & 1.07 & $<0.5$\\
1324+0320 & $-0.47\pm{0.22}$ & 1.82 & 8.0 (1.0-20.0)\\
1328+5818 & $0.00\pm{0.21}$  & 1.58 & 40.0 (20.0-50.0)\\
1336+5146 & $-0.79\pm{0.18}$ & 0.92 & $<1.0$\\  
1340-0019 & $-0.75\pm{0.15}$ & 1.79 & $<1.0$ \\
1425+3739 & $-0.38\pm{0.18}$ & 1.04 & 8.0 (3.0-15.0)\\
1426+3753 & $-0.62\pm{0.22}$ & 1.10 & 1.0 (0.05-2.0)\\
1435+4841 & $-0.50\pm{0.20}$ & 2.91 & 2.0 (0.01-8.0)\\
1446+0255 & $-0.25\pm{0.20}$ & 3.16 & 6.0 (3.0-10.0)\\
1508+5652 & $-0.62\pm{0.16}$ & 1.64 & 1.0 (0.01-3.0)\\
1517+0016 & $<-0.72$         & 4.36 & $<0.1$ \\
1533+3243 & $-0.53\pm{0.12}$ & 2.24 & 2.0 (0.05-3.0)\\
2039-0102 & $-0.64\pm{0.34}$ & 6.30 & 1.0 (0.01-6.0)\\ 
\hline
\end{tabular}\\
\begin{flushleft}
\small{Notes: Col.(1): Source name; Col.(2): Hardness ratio computed between (0.2-2) and (2-8) keV. Errors are at 1$\sigma$, propagated from the count-rate errors using the numerical method of \citet{lyons}; Col.(3): Galactic neutral hydrogen column density, taken from \citet{1990ARA&A..28..215D}; Col.(4): Neutral hydrogen column density at the source redshift estimated from hardness ratio, where the number in parentheses are the 1$\sigma$ confidence intervals.}
\end{flushleft}
\end{table*}

\begin{table*}
\caption{Optical/X properties}
\label{table:4}
\centering          
\begin{tabular}{ c c c c c c c c c c}
\hline\hline
Name & $\log f_{\rm{2\,keV}}$ & $\log f_{\rm{2500}}$  & $\alpha_{\rm{ox}}$ & $\log l_{\rm{2500}}$ & $\alpha_{\rm{ox}}\left(l_{\rm{2500}}\right)$ &   $\Delta\alpha_{\rm{ox}}$ & Type & AI & $v_{\rm{out}}$  \\
(1) & (2) & (3) & (4) & (5) & (6) & (7) & (8) & (9) & (10) \\
\hline
0042-0912 & -31.26  & -27.06 & -1.61 & 30.84 & -1.61 & 0.00  & Hi & 276 & 9400 \\
0231-0739 & -31.11  & -27.38 & -1.43 & 30.77 & -1.60 & 0.17  & H  & 510 & 19400\\
0243+0000 & -31.24  & -26.68 & -1.75 & 31.31 & -1.68 & -0.07 & Hi & 1792 & 6200 \\
0734+3203 & -31.67  & -27.31 & -1.67 & 30.71 & -1.59 & -0.08 & Hi & 333 & 29000 \\
0855+3757 & -31.40  & -26.77 & -1.78 & 31.18 & -1.66 & -0.12 & Hi & 1482 & 21200\\
0911+0550 & -30.67  & -26.55 & -1.58 & 31.69 & -1.73 & 0.15  & H  & 1149 & 21500\\
0921+5131 & -31.12  & -27.00 & -1.58 & 30.92 & -1.62 & 0.04  & Hi & 404 & 1600 \\ 
0922+5121 & -31.16  & -27.50 & -1.40 & 30.38 & -1.55 & 0.15  & Lo & 1075 & 9600 \\
0923+5127 & -31.19  & -27.04 & -1.59 & 31.00 & -1.63 & 0.04  & Hi & 3956 & 19400 \\
1001+0140 & -31.23  & -27.28 & -1.52 & 30.72 & -1.59 & 0.07  & Hi & 427 & 5000 \\
1001+0224 & -31.47  & -27.47 & -1.54 & 30.53 & -1.57 & 0.03  & Hi & 547 & 1100 \\
1007+5343 & -31.23  & -27.09 & -1.59 & 30.80 & -1.61 & 0.02  & Lo & 748 & 6000 \\
1011+5541 & -31.66  & -27.38 & -1.64 & 30.85 & -1.61 & -0.03 & H  & 2139 & 4100 \\
1019+0825 & -30.78  & -26.47 & -1.66 & 31.82 & -1.75 & 0.09  & H & 425 & 26900 \\
1106+5222 & -31.93  & -27.05 & -1.87 & 30.87 & -1.62 & -0.25 & Hi & 2048 & 3900 \\
1123+0524 & -31.48  & -27.30 & -1.60 & 31.13 & -1.65 & 0.05  & H & 508 & 8800 \\
1135+4913 & -31.72  & -27.01 & -1.81 & 30.97 & -1.63 & -0.18 & Hi & 306 & 13700 \\ 
1141-0143 & -31.78  & -26.55 & -2.01 & 31.06 & -1.64 & -0.37 & Lo & 261 & 4100 \\
1223+1034 & -30.99  & -26.93 & -1.56 & 31.29 & -1.68 & 0.12  & H  & 1707 & 5300 \\
1227+0126 & -32.09  & -27.05 & -1.93 & 30.92 & -1.62 & -0.31 & Hi & 1820 & 18100 \\
1236+6158 & -31.77  & -26.79 & -1.91 & 31.37 & -1.69 & -0.22 & H & 621 & 14100 \\
1245-0021 & -31.57  & -27.34 & -1.62 & 30.76 & -1.60 & -0.02 & H  & 2735 & 4400 \\
1324+0320 & -31.80  & -26.89 & -1.89 & 31.48 & -1.70 & -0.19 & H & 1369 & 3200 \\
1328+5818 & -31.63  & -26.75 & -1.87 & 31.57 & -1.71 & -0.16 & H & 3890 & 8000 \\
1340-0019 & -31.40  & -27.32 & -1.56 & 30.60 & -1.58 & 0.02  & Hi & 488 & 2560 \\
1435+4841 & -31.09  & -26.99 & -1.57 & 30.94 & -1.63 & 0.06  & Hi & 1879 & 3800 \\
1446+0255 & -31.72  & -27.03 & -1.80 & 30.90 & -1.62 & -0.18 & Hi & 1784 & 6900 \\
1508+5652 & -31.16  & -26.99 & -1.60 & 30.91 & -1.62 & 0.02  & Hi & 979 & 2900 \\
1517+0016 & -31.86  & -26.93 & -1.89 & 31.01 & -1.64 & -0.25 & Hi & 451 & 14100 \\
1525+5136 & -30.39  & -26.09 & -1.65 & 32.17 & -1.80 & 0.15  & H  & 6216 & 28900 \\
1533+3243 & -31.26  & -26.94 & -1.66 & 31.01 & -1.64 & -0.02 & Hi & 714 & 4100 \\
1543+5359 & -30.67  & -26.19 & -1.72 & 31.92 & -1.76 & 0.04  & H  & 503 & 14000 \\
2039-0102 & -31.56  & -26.90 & -1.79 & 31.11 & -1.65 & -0.14 & Hi & 311 & 4900 \\
 & & & & & \\
\hline
 & & & & & \\
0043+0046 & -31.04  & -27.31 & -1.43 & 30.48 & -1.56 & 0.13  & Lo & 530(*) & 15800)(*)\\
0043+0052 & -31.25  & -27.99 & -1.25 & 29.27 & -1.39 & 0.14  & Lo & 570(*) & 17000(*)\\
0109+1328 & -31.40  & -27.07 & -1.66 & 30.52 & -1.57 & -0.09 & Lo & 180(*) & 1000(*) \\
0242-0000 & -31.75  & -26.98 & -1.83 & 31.17 & -1.66 & -0.17 & H & 6700(*) &  12000(*)      \\
0302+0006 & -31.49  & -27.48 & -1.54 & 30.88 & -1.62 & 0.08  & H  & 1570(*) & 4000(*)\\
0729+3700 & -31.60  & -27.04 & -1.75 & 30.35 & -1.54 & -0.21 & Lo & 430(*) & 25000(*) \\
0919+3030 & -31.68  & -26.85 & -1.86 & 30.84 & -1.61 & -0.25 & Lo & 215(*) & 25000(*) \\
0939+3556 & -31.89  & -26.86 & -1.93 & 31.14 & -1.65 & -0.28 & Hi & 520(*) & 16000(*) \\
0943+4811 & -31.78  & -26.96 & -1.85 & 30.95 & -1.63 & -0.22 & Hi & 640(*) & 20000(*) \\
1001+5553 & -29.54  & -26.18 & -1.29 & 31.52 & -1.71 & 0.42  & Lo & 150(*) & 25000(*)\\
1021+1315 & -30.70  & -26.78 & -1.50 & 31.01 & -1.64 & 0.14  & Lo & 550(*) & 25000(*)\\
1052+4414 & -31.53  & -26.91 & -1.77 & 30.98 & -1.63 & -0.14 & Hi & 4500(*) & 25000(*) \\
1124+3851 & -31.69  & -27.33 & -1.67 & 31.07 & -1.64 & -0.03 & H & 1320(*) & 14500(*) \\
1146+4723 & -30.57  & -26.97 & -1.38 & 30.97 & -1.63 & 0.25  & Lo & 550(*) & 12600(*)\\
1205+4431 & -31.60  & -26.80 & -1.84 & 31.15 & -1.66 & -0.18 & Hi & 1900(*) & 20000(*)\\
1335+5147 & -31.52  & -26.70 & -1.85 & 31.22 & -1.67 & -0.18 & Lo & 1890(*) & 9000(*)\\
1336+5146 & -31.93  & -26.99 & -1.90 & 31.08 & -1.65 & -0.25 & H & 4300(*)  & 7600(*) \\ 
1425+3739 & -31.65  & -27.00 & -1.78 & 31.21 & -1.66 & -0.12 & H & 550(*) & 20000(*) \\
1425+3757 & -30.95  & -26.69 & -1.63 & 31.26 & -1.67 & 0.04  & Hi & 1880(*) & 25000(*)\\
1426+3753 & -31.56  & -27.10 & -1.71 & 30.80 & -1.61 & -0.10 & Hi & 790(*) & 25000(*) \\ 
1447+4032 & -30.94  & -26.75 & -1.61 & 30.91 & -1.62 & 0.01  & Lo & 670(*) & 17000(*)\\ 
\hline
\end{tabular}
\begin{flushleft}
\small{Notes: Col.(1): Source name; Col.(2): Logarithm of 2~keV observed flux density in units of erg s$^{-1}$ cm$^{-2}$ Hz$^{-1}$; Col.(3): Logarithm of $2500\:\AA$ observed flux density in units of erg s$^{-1}$ cm$^{-2}$ Hz$^{-1}$; Col.(4): Observed optical-X spectral index; Col.(5): Logarithm of $2500\:\AA$ intrinsic luminosity density in units of erg s$^{-1}$ Hz$^{-1}$; Col.(6): Optical-X spectral index expected from Eq.(3) of \citet{2007ApJ...665.1004J}; Col.(7): Difference between observed and expected $\alpha_{\rm{ox}}$; Col.(8): BALQSO subclassification, Lo=LoBAL, Hi=HiBAL, H=BALQSO for which the Mg~II spectral region is redshifted outside the SDSS window, so preventing from checking the presence of low-ionization absorption troughs; Col.(9): Absorption Index (AI) taken from  \citet{2006ApJS..165....1T} when available, in units of km s$^{-1}$; Col.(10): Maximum velocity outflow taken from  \citet{2006ApJS..165....1T} when available, in units of km s$^{-1}$. The values of AI and $v_{\rm{out}}$ marked with (*) are estimated by us from SDSS spectra. }
\end{flushleft}
\end{table*}

\end{appendix}
\clearpage
\begin{appendix}
\section{Source Spectra}\label{appendix:1}

\begin{figure*}
\centering
\subfigure{
\includegraphics[width=8cm]{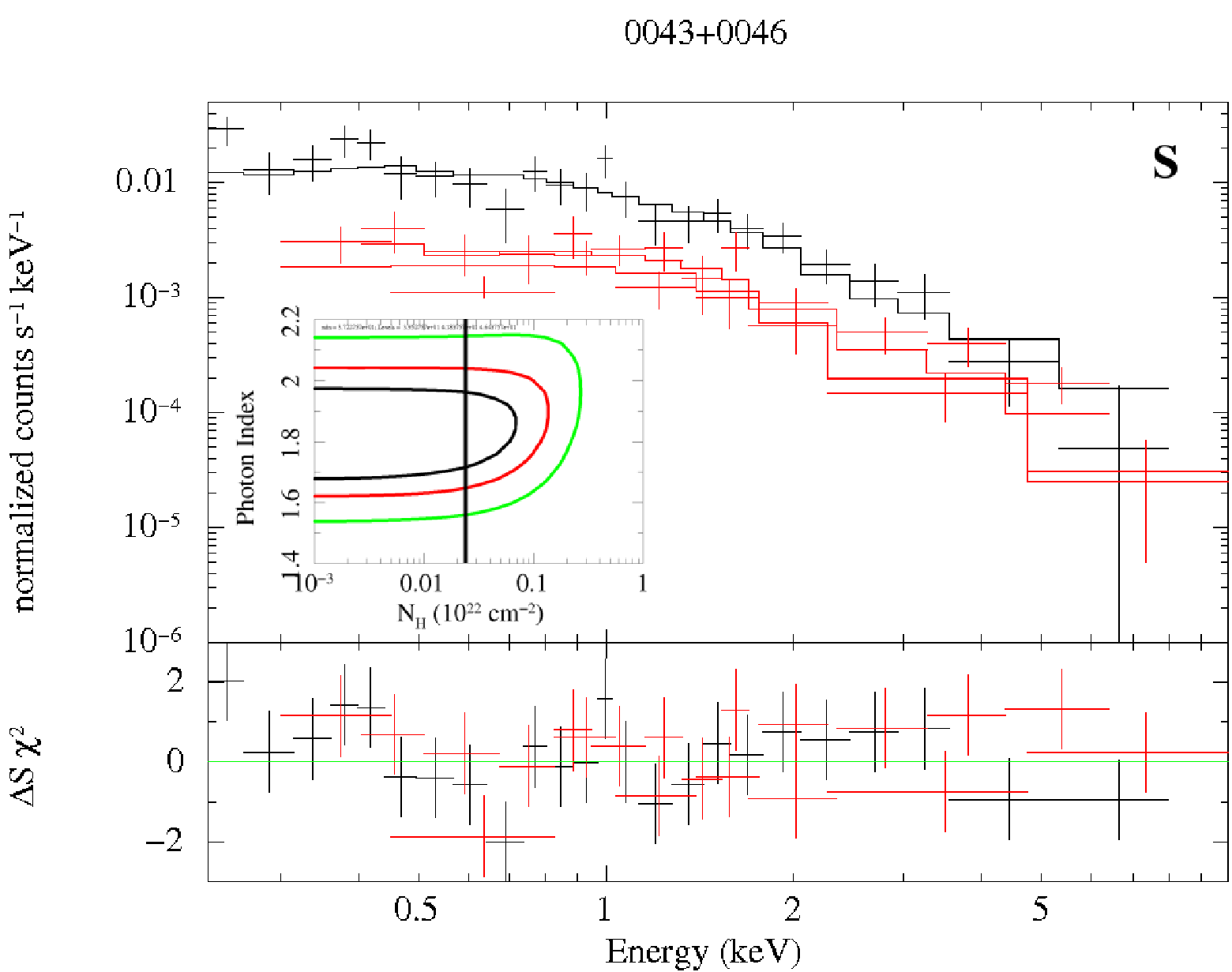}\hspace{2cm}
\includegraphics[width=8cm]{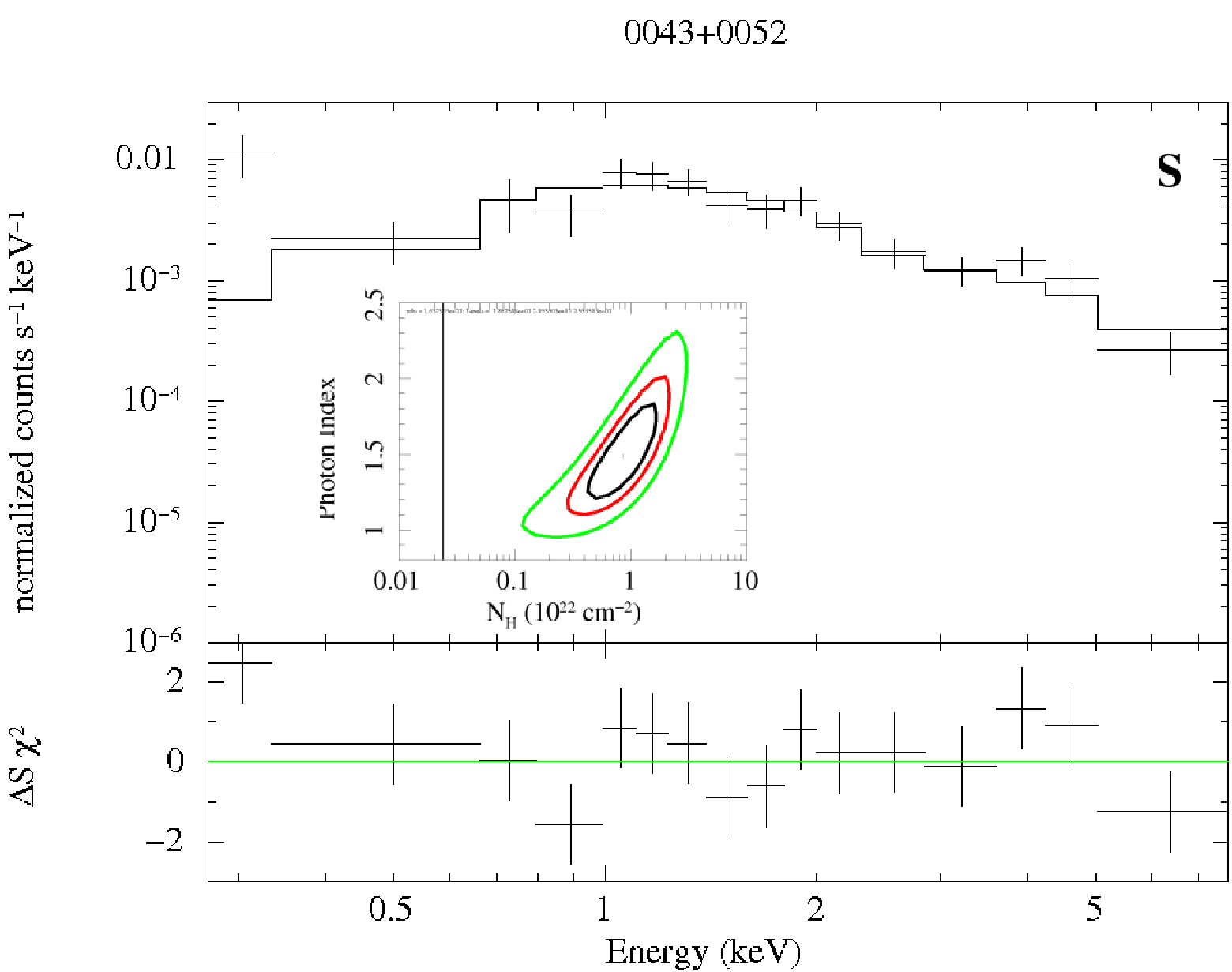}} 
\centering
\subfigure{
\includegraphics[width=8cm]{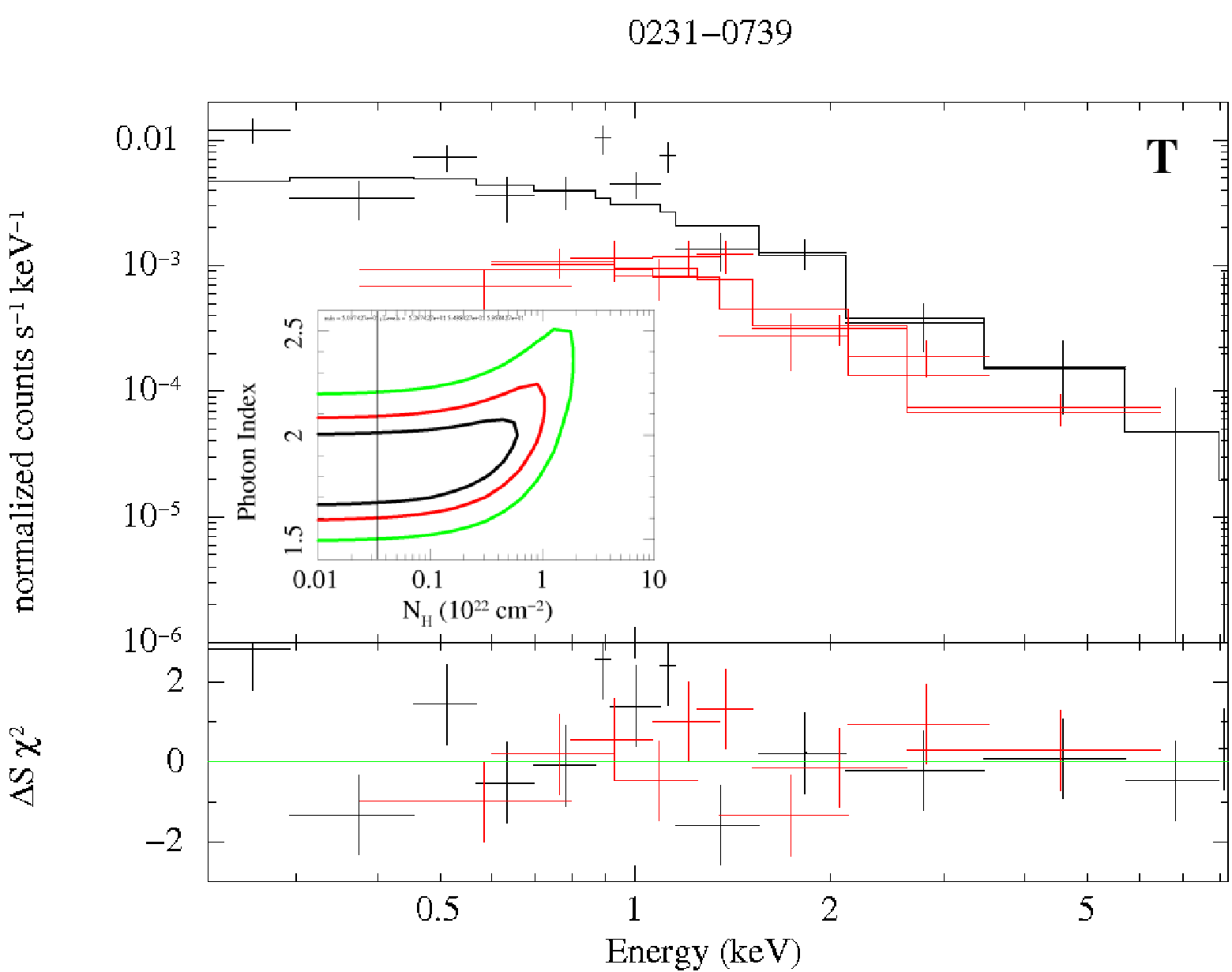}
\hspace{2cm}
\includegraphics[width=8cm]{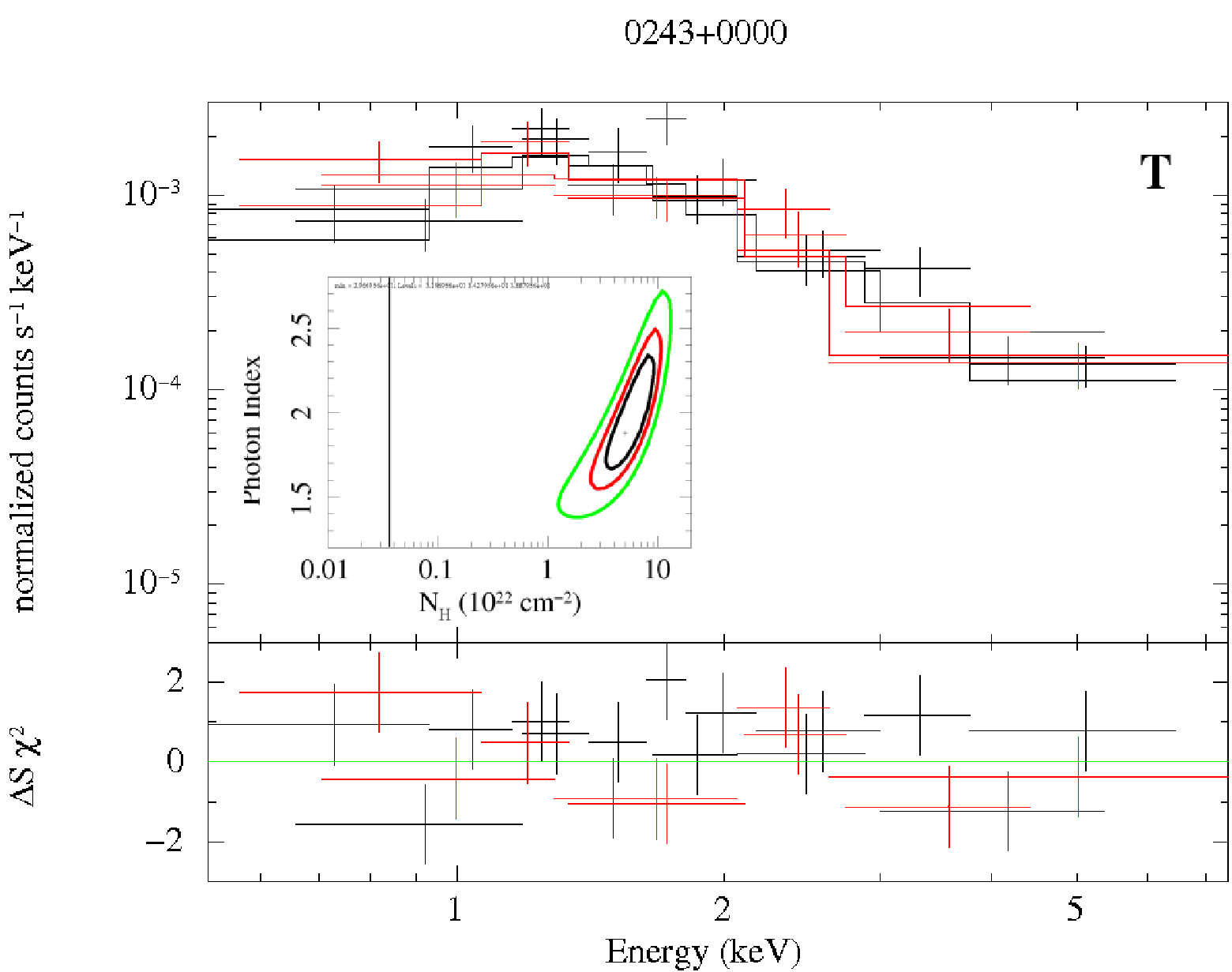}
}
\centering
\subfigure{
\includegraphics[width=8cm]{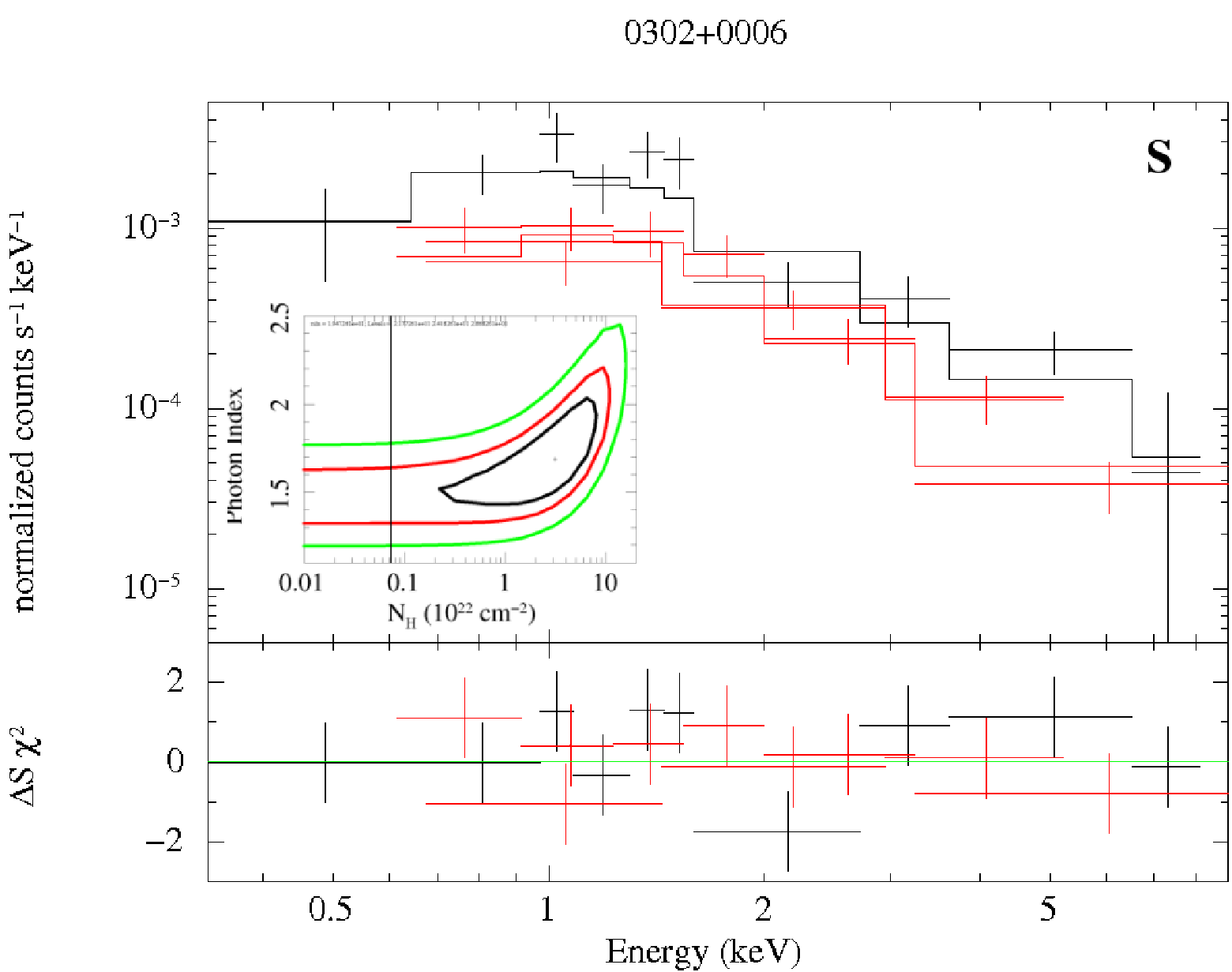}
\hspace{2cm}
\includegraphics[width=8cm]{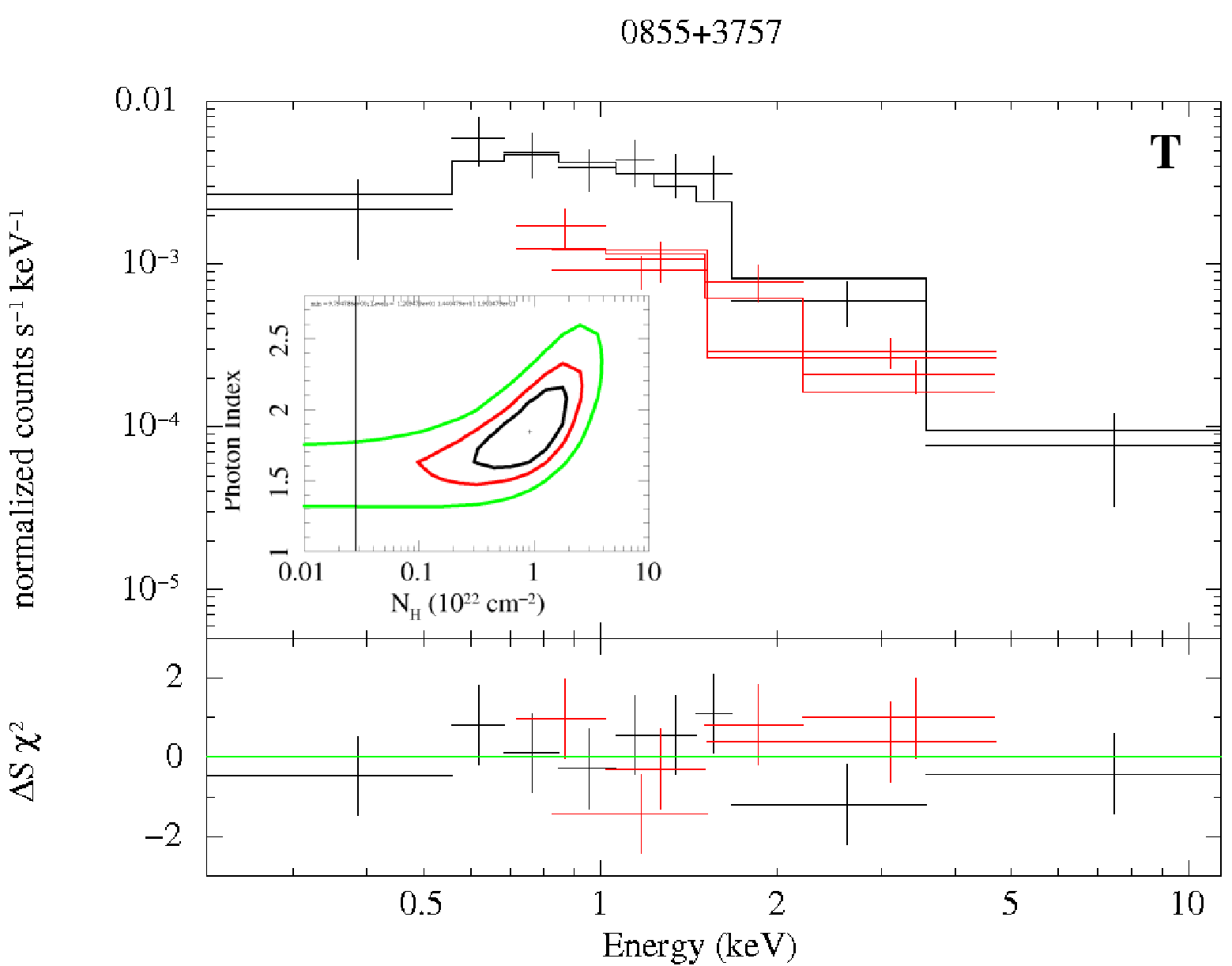}
}
\caption{\label{fig:graphics:a}Source spectra of the 22 BALQSOs with spectral analysis available. EPIC-pn data points are thick (black in the on-line version), while EPIC-MOS data are thin (red in the on-line version). The inset contains confidence contours at 68, 90 and 99\% confidence level for column density $N_{\rm{H}}$ and photon index $\Gamma$, where the vertical line marks the amount of Galactic neutral absorption. In the top-right corner is reported a ``T'' for BALQSOs selected from \citet{2006ApJS..165....1T}, a ``S'' for BALQSOs selected from \citet{2008ApJ...680..169S}.}
\label{fig:graphics}
\end{figure*}
\addtocounter{figure}{-1}
\begin{figure*}
\addtocounter{subfigure}{1}
\centering
\subfigure{\includegraphics[width=8cm]{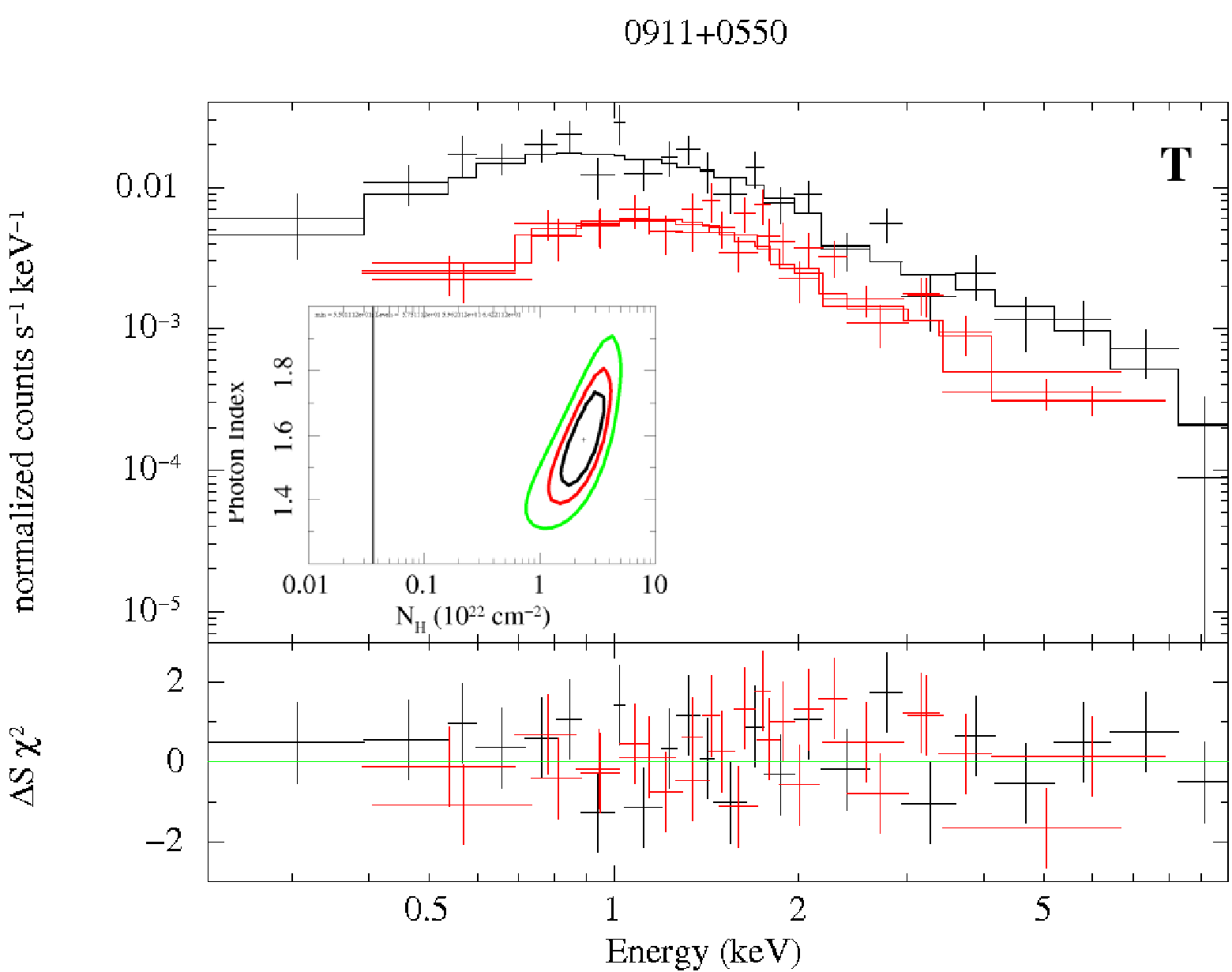}\hspace{2cm}
\includegraphics[width=8cm]{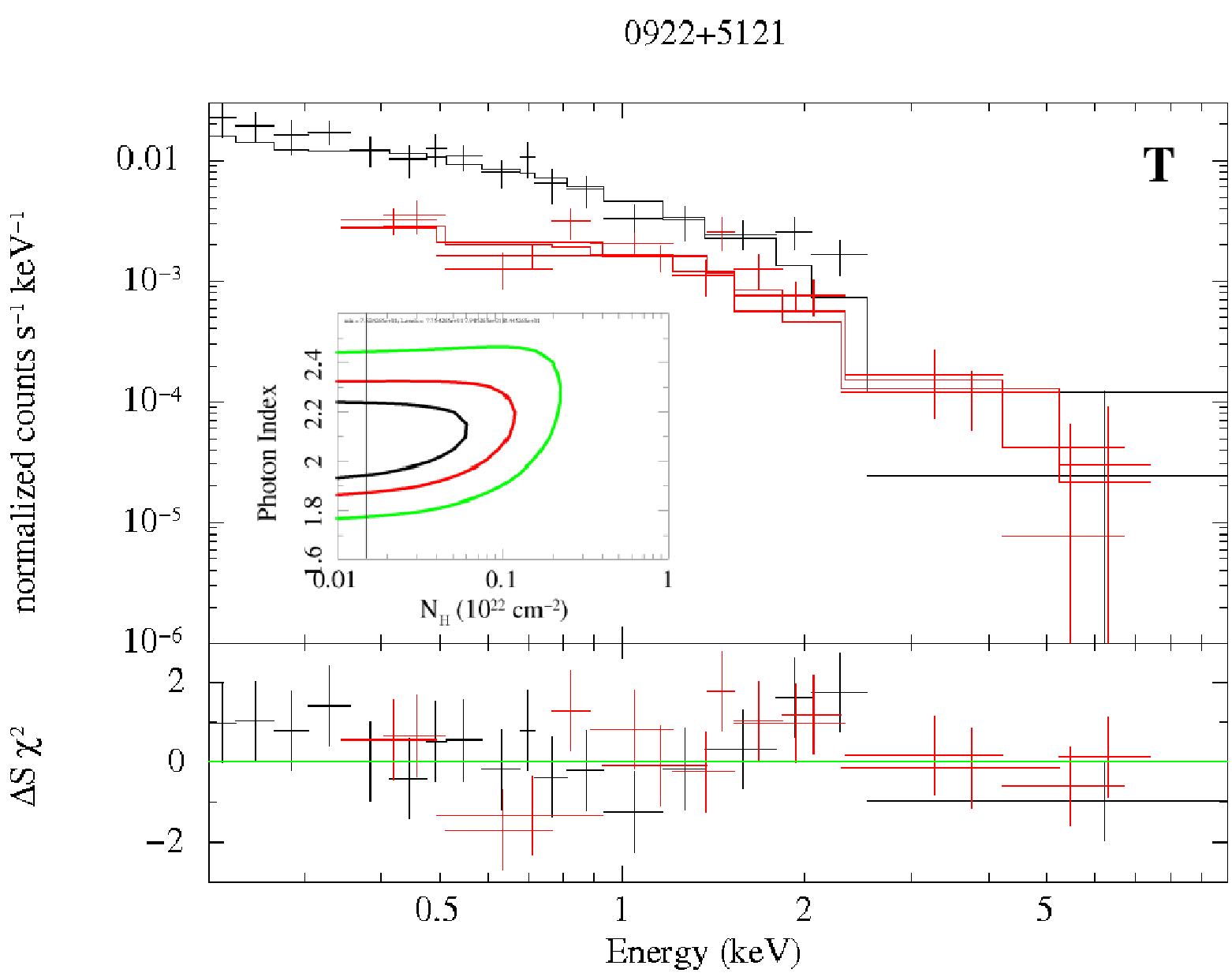}}\\
\subfigure{
\includegraphics[width=8cm]{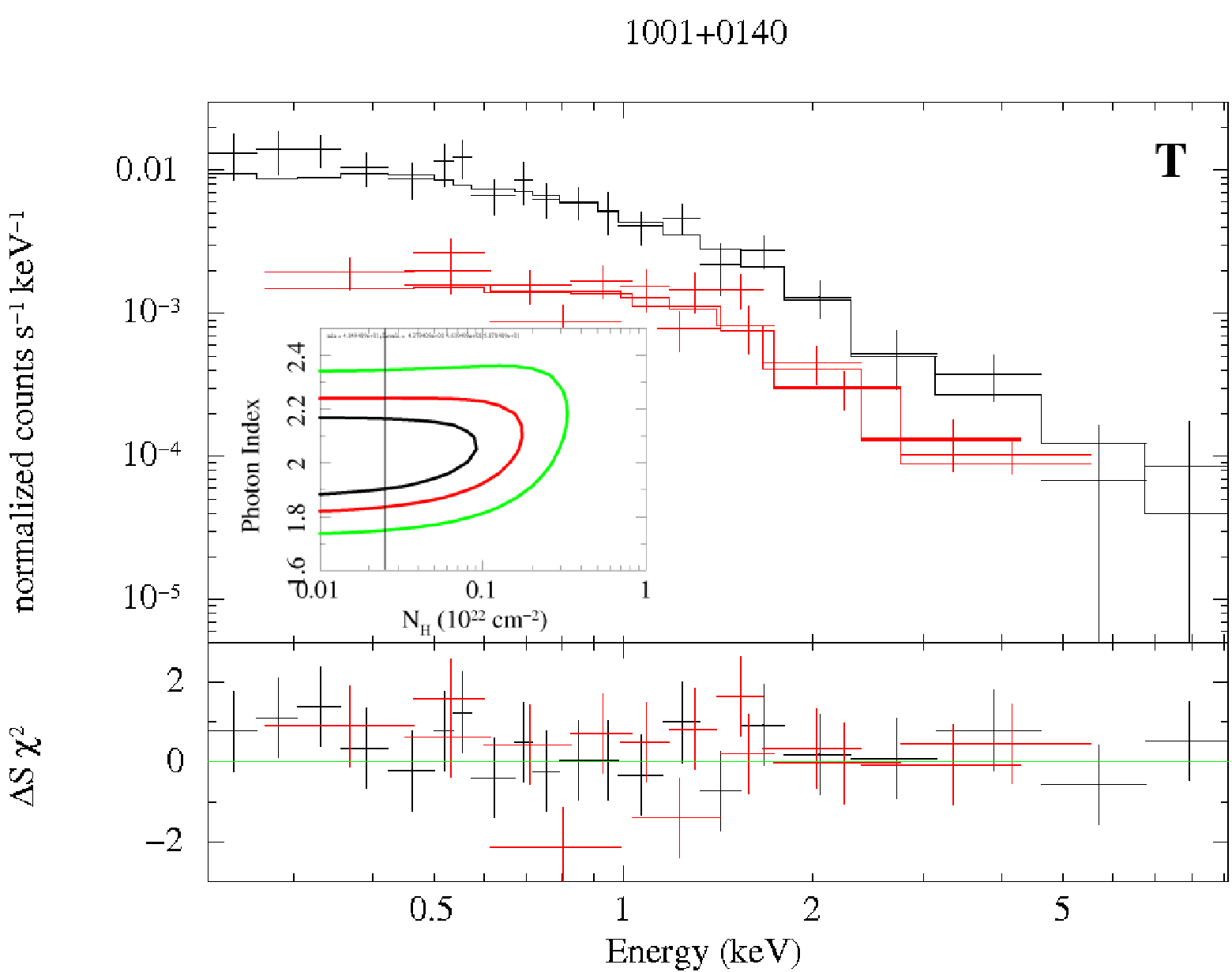}\hspace{2cm}
\includegraphics[width=8cm]{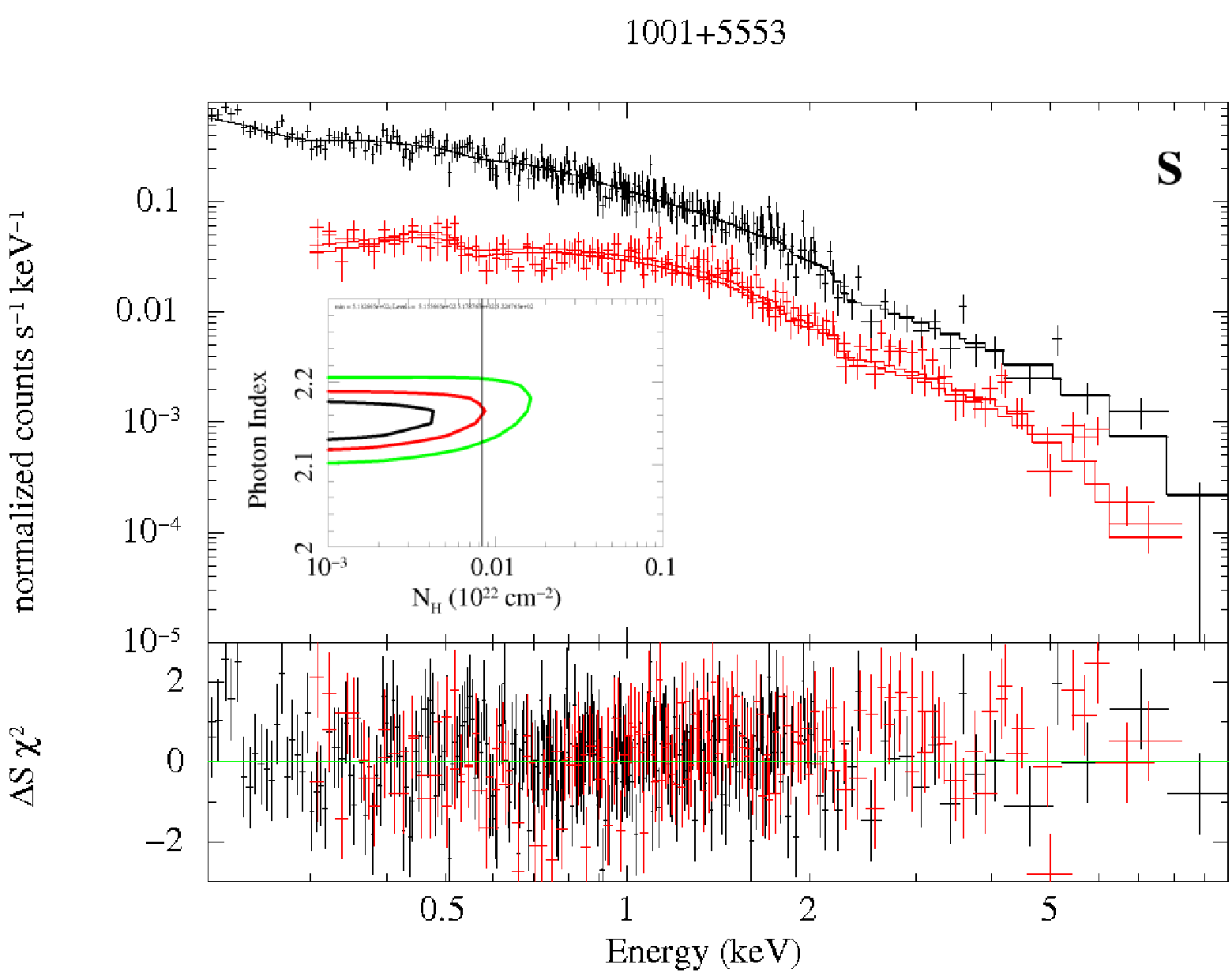}}\\
\subfigure{
\includegraphics[width=8cm]{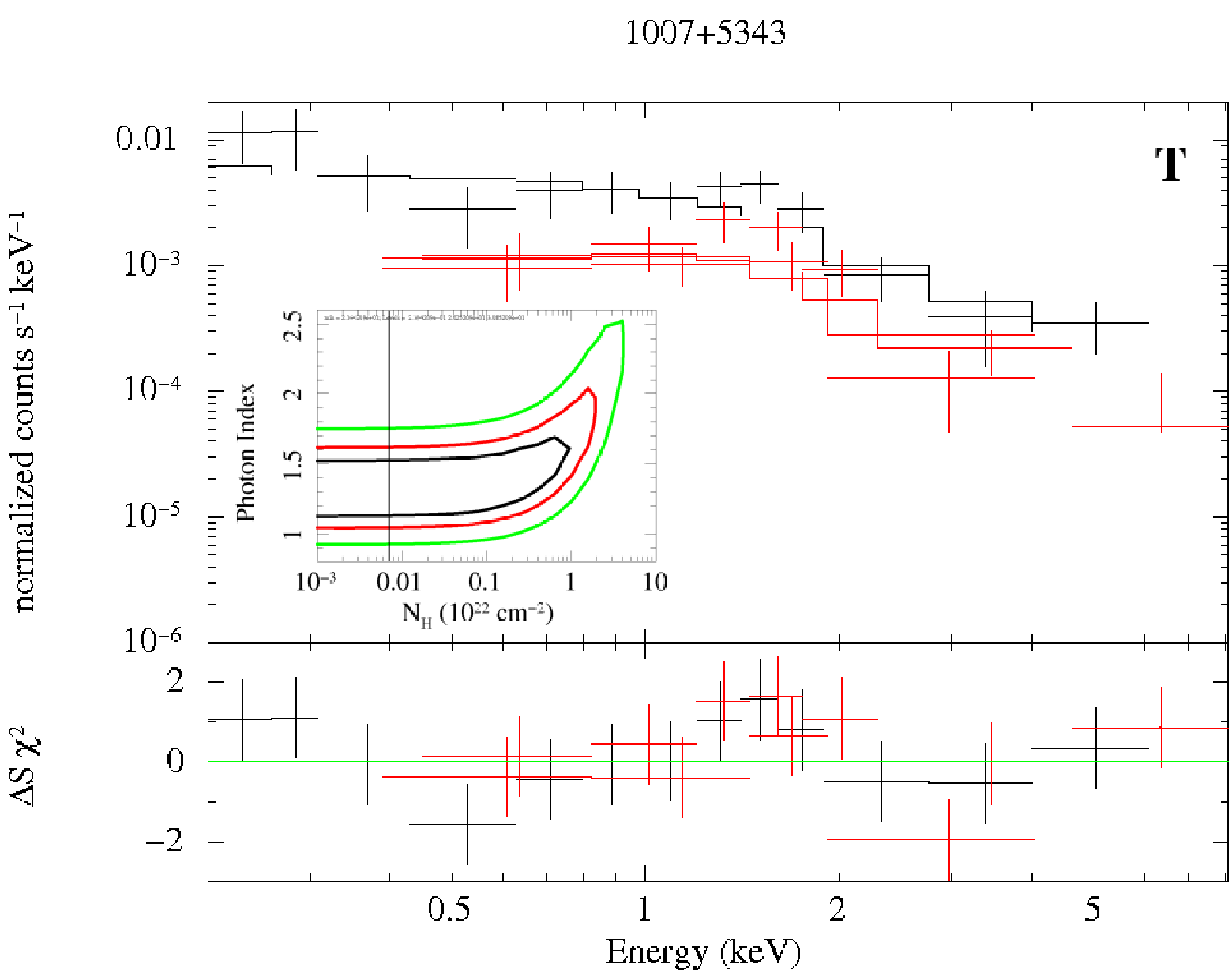}\hspace{2cm}
\includegraphics[width=8cm]{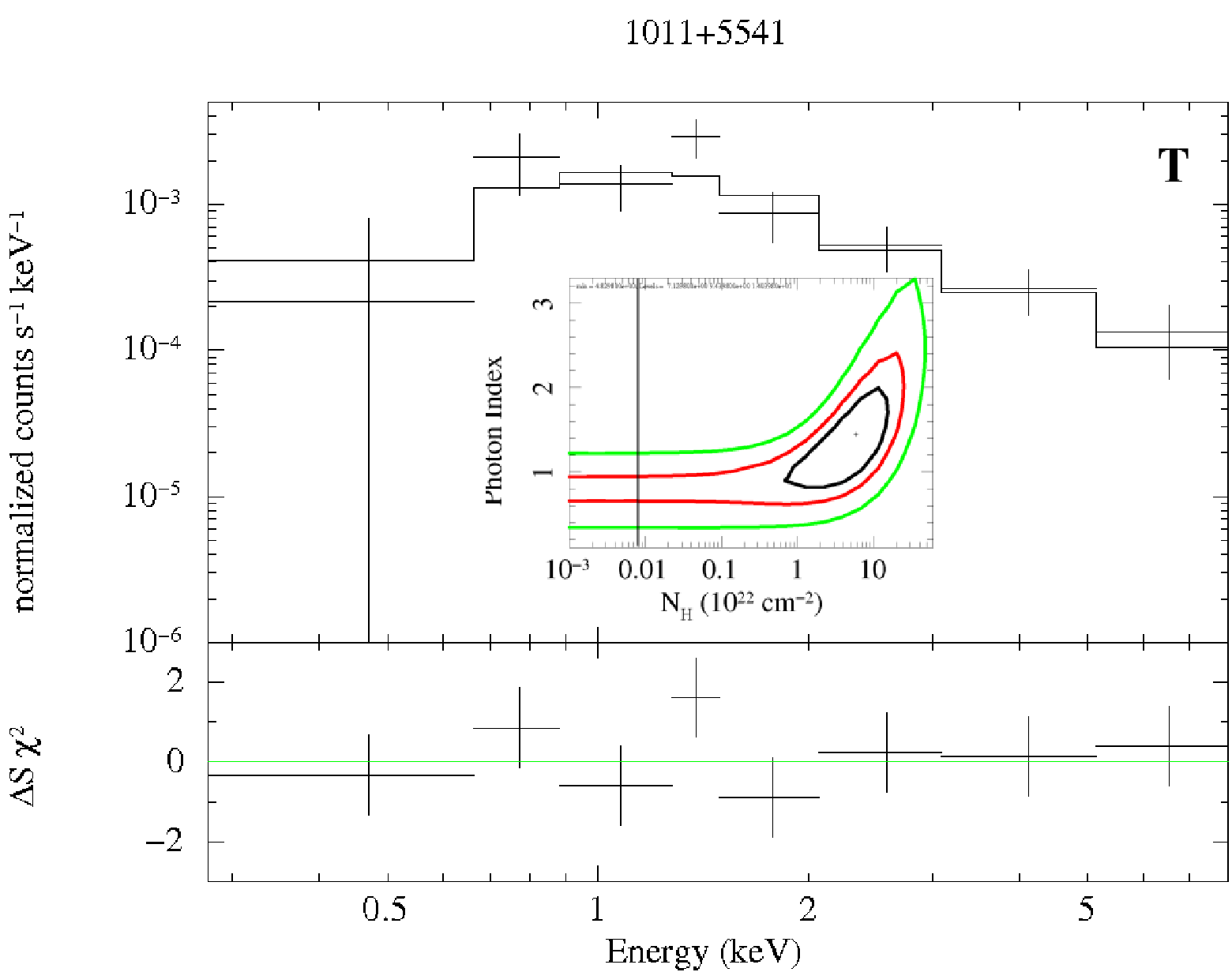}
 }%
\caption{ -- Continued}%
\end{figure*}

\addtocounter{figure}{-1}
\begin{figure*}
\addtocounter{subfigure}{1}
\centering
\subfigure{\includegraphics[width=8cm]{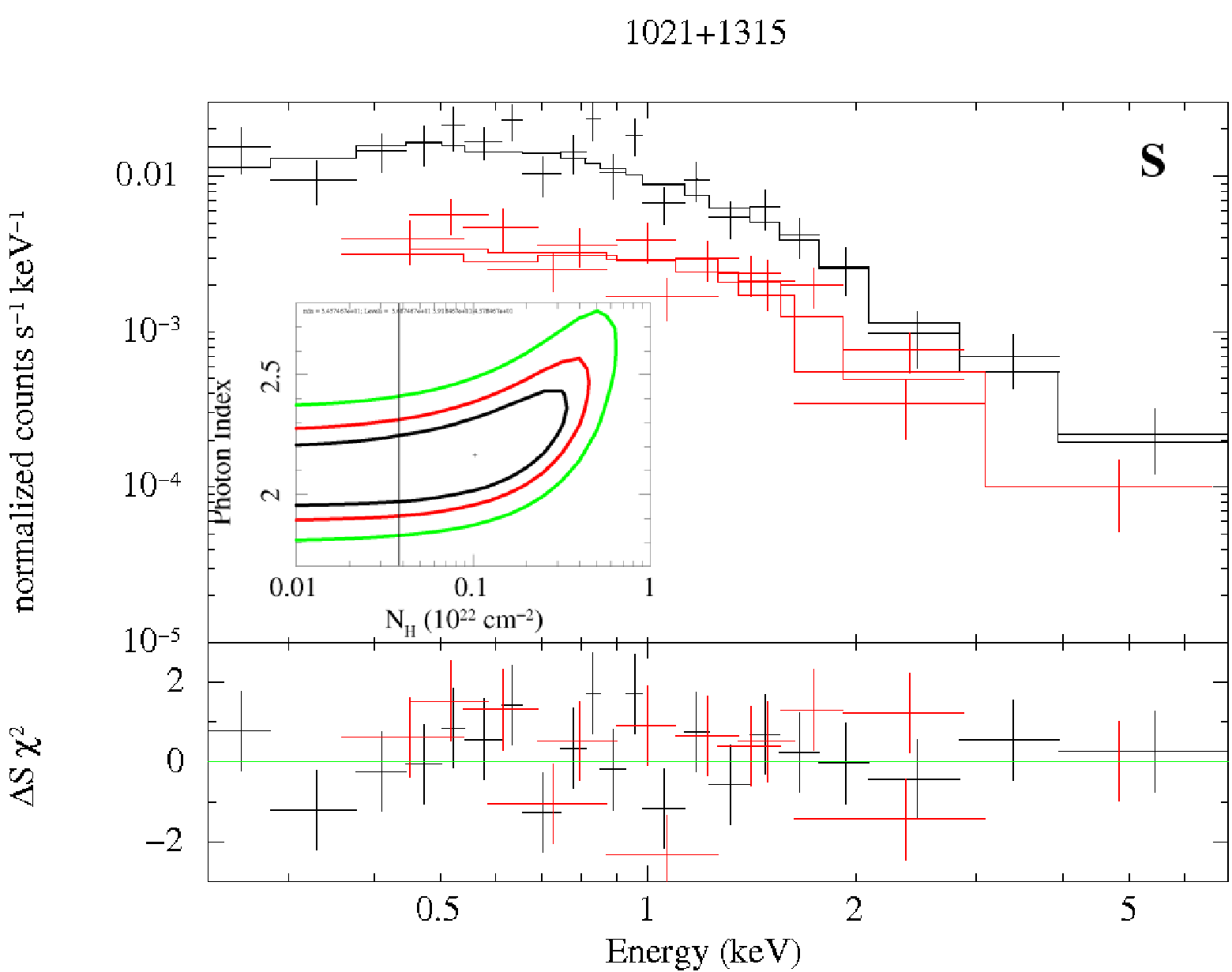}\hspace{2cm}
\includegraphics[width=8cm]{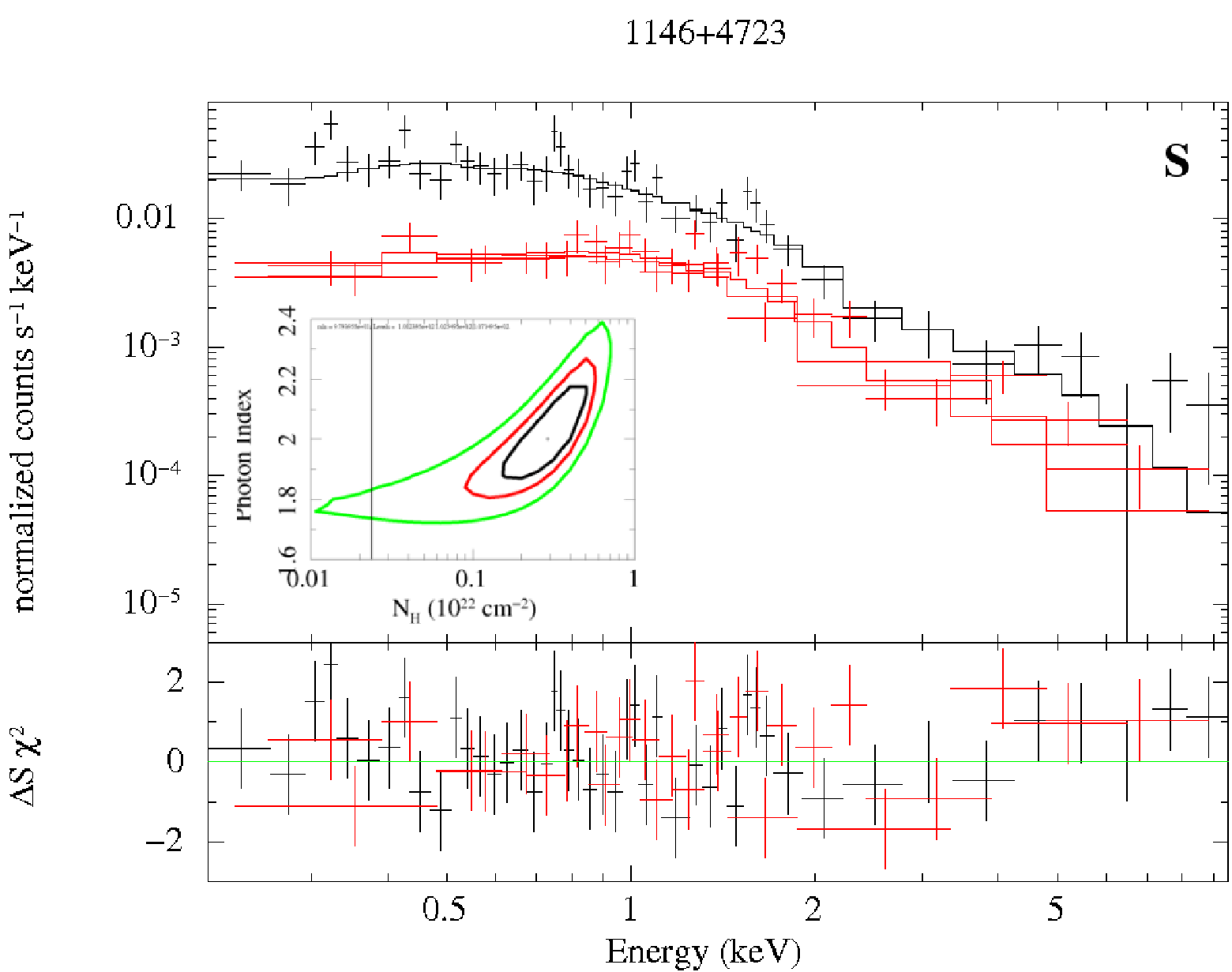}}\\
\subfigure{
\includegraphics[width=8cm]{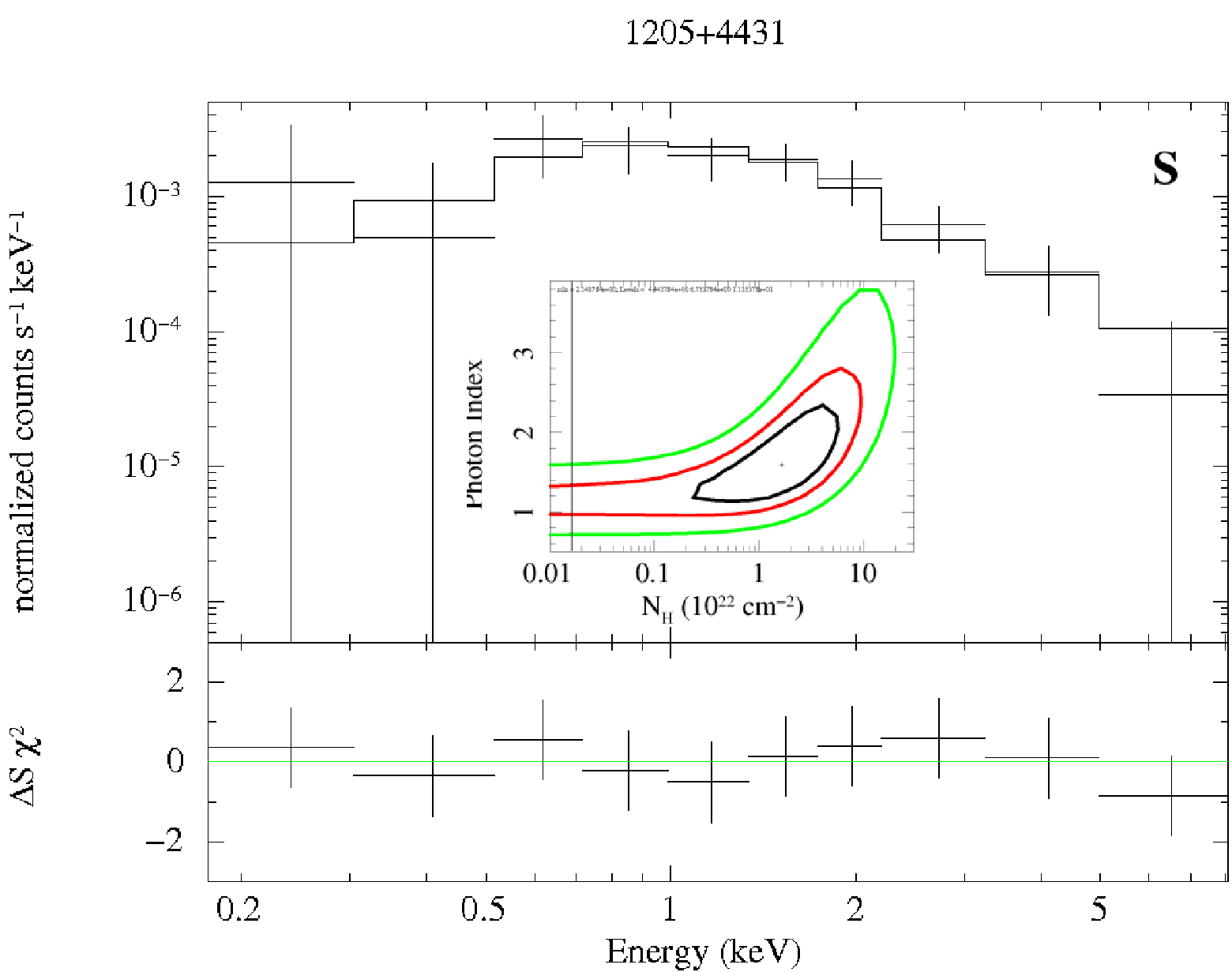}\hspace{2cm}
\includegraphics[width=8cm]{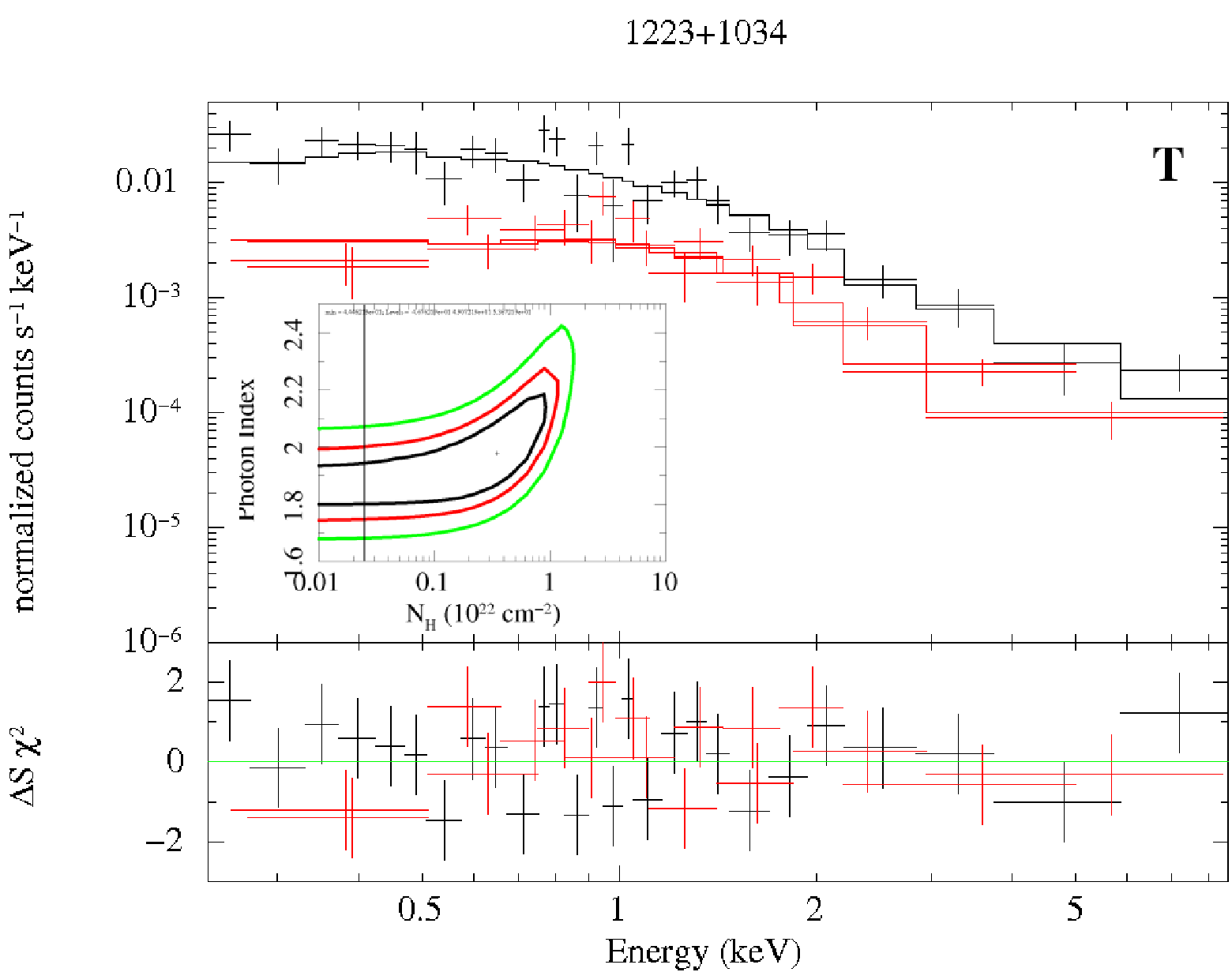}}\\
\subfigure{
\includegraphics[width=8cm]{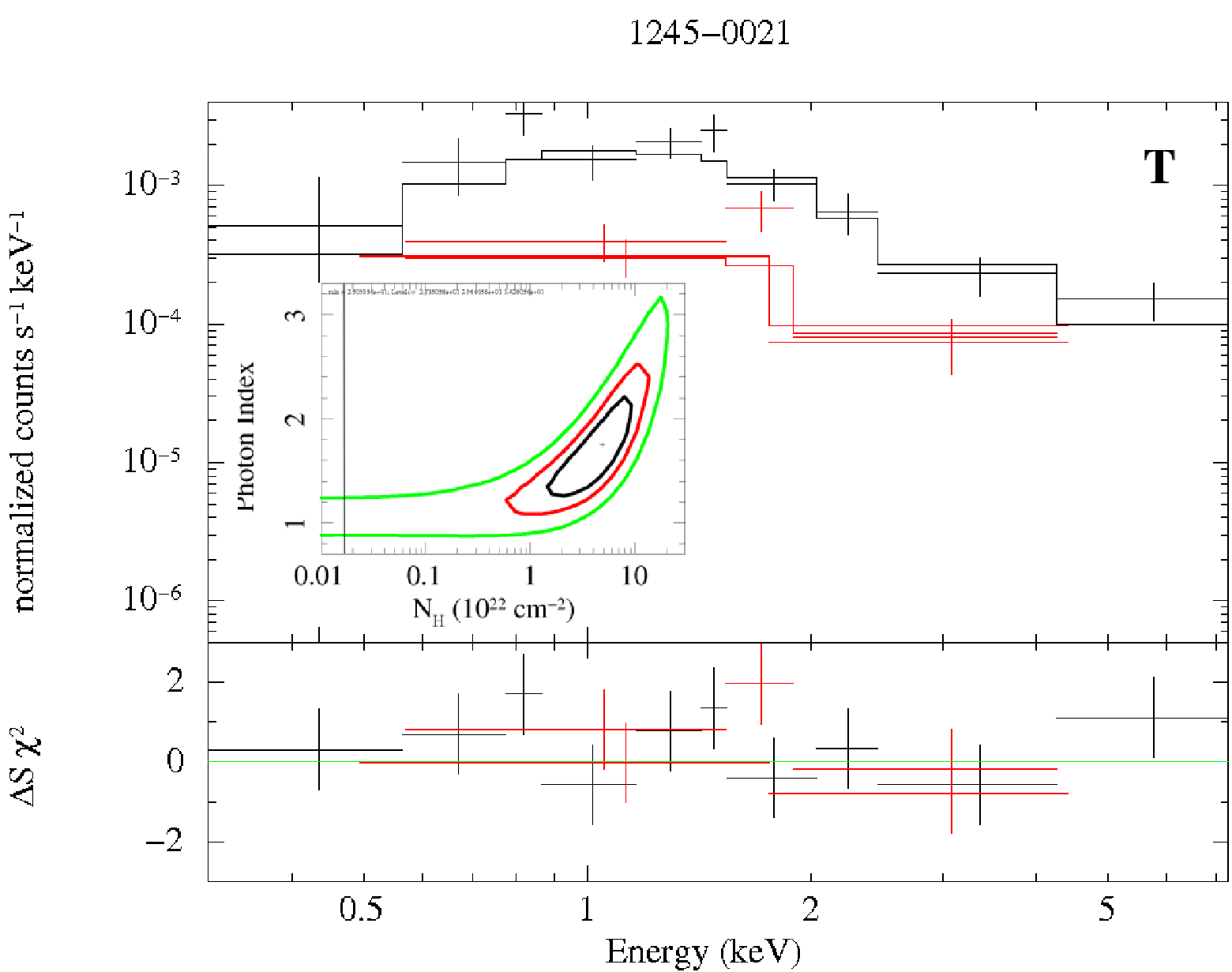}\hspace{2cm}
\includegraphics[width=8cm]{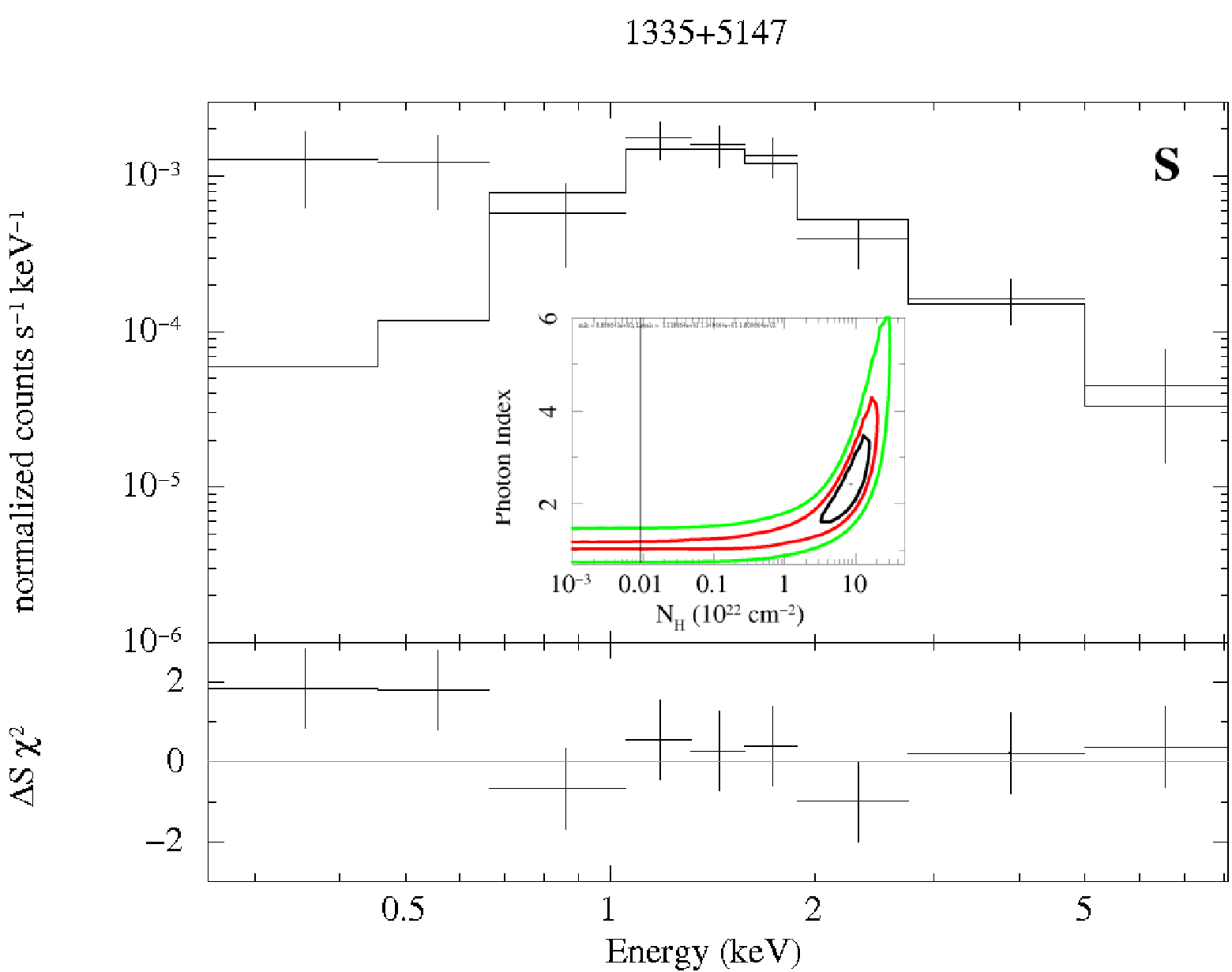}
 }%
\caption{ -- Continued}%
\end{figure*}

\addtocounter{figure}{-1}
\begin{figure*}
\addtocounter{subfigure}{1}
\centering
\subfigure{\includegraphics[width=8cm]{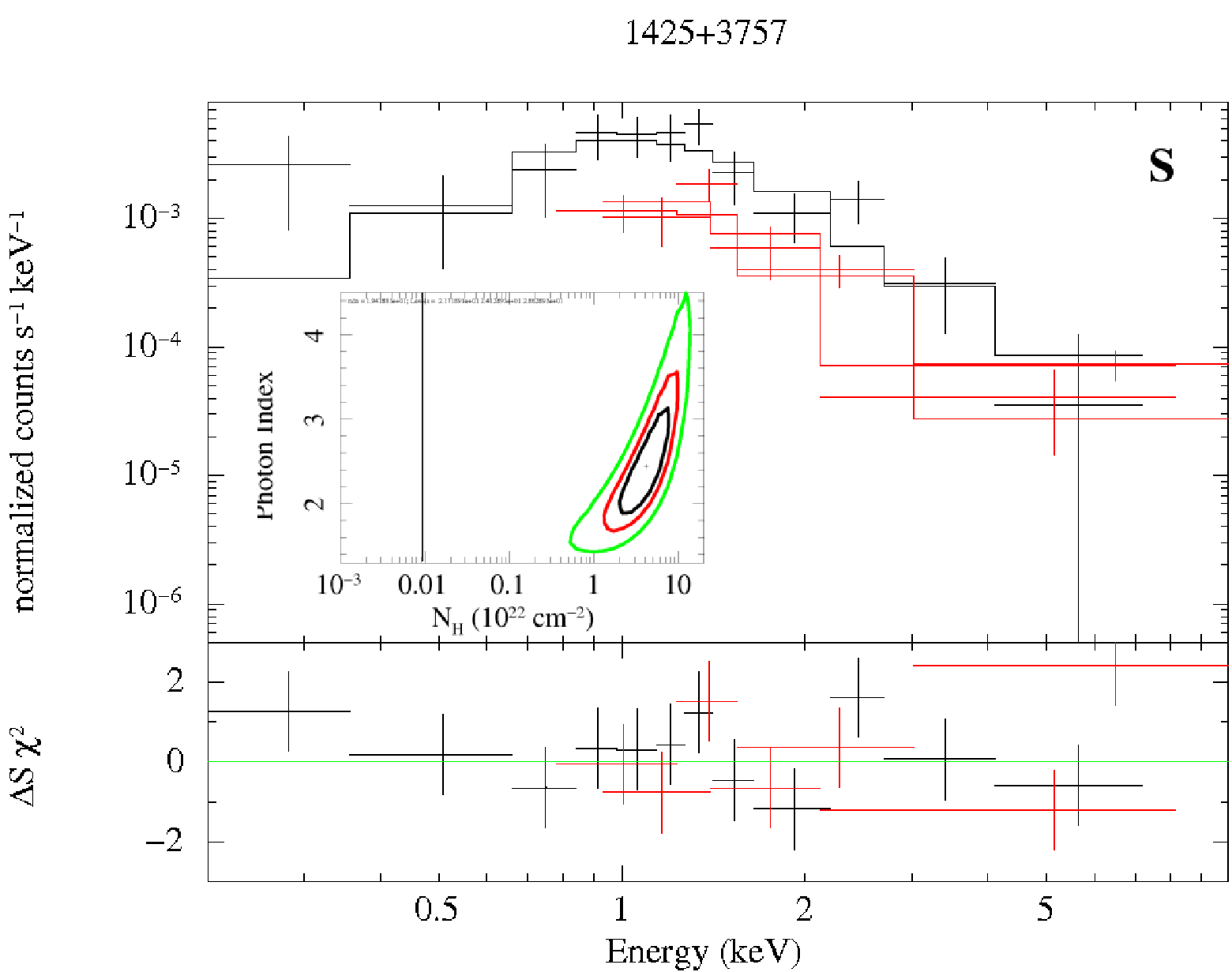}\hspace{2cm}
\includegraphics[width=8cm]{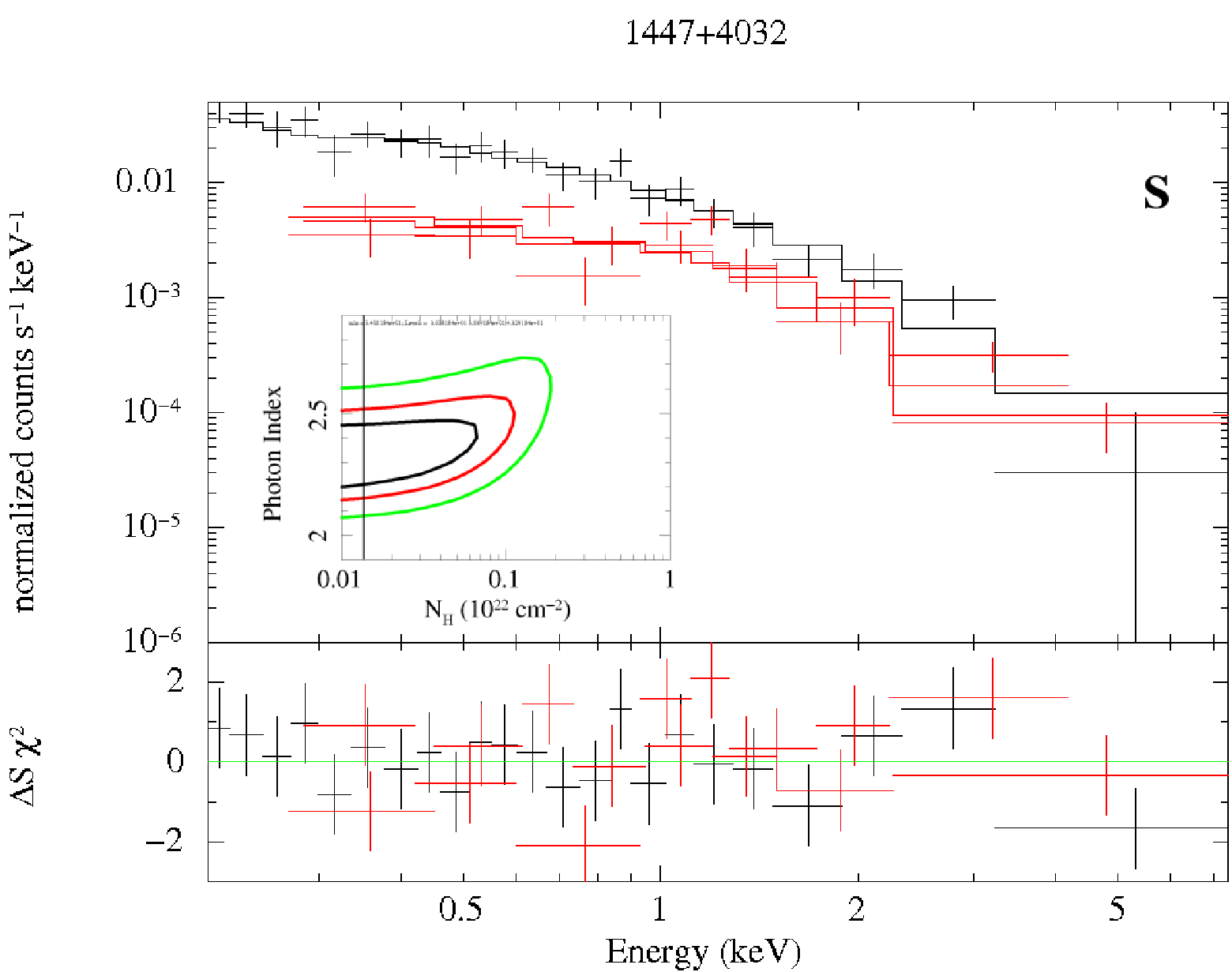}}\\
\subfigure{
\includegraphics[width=8cm]{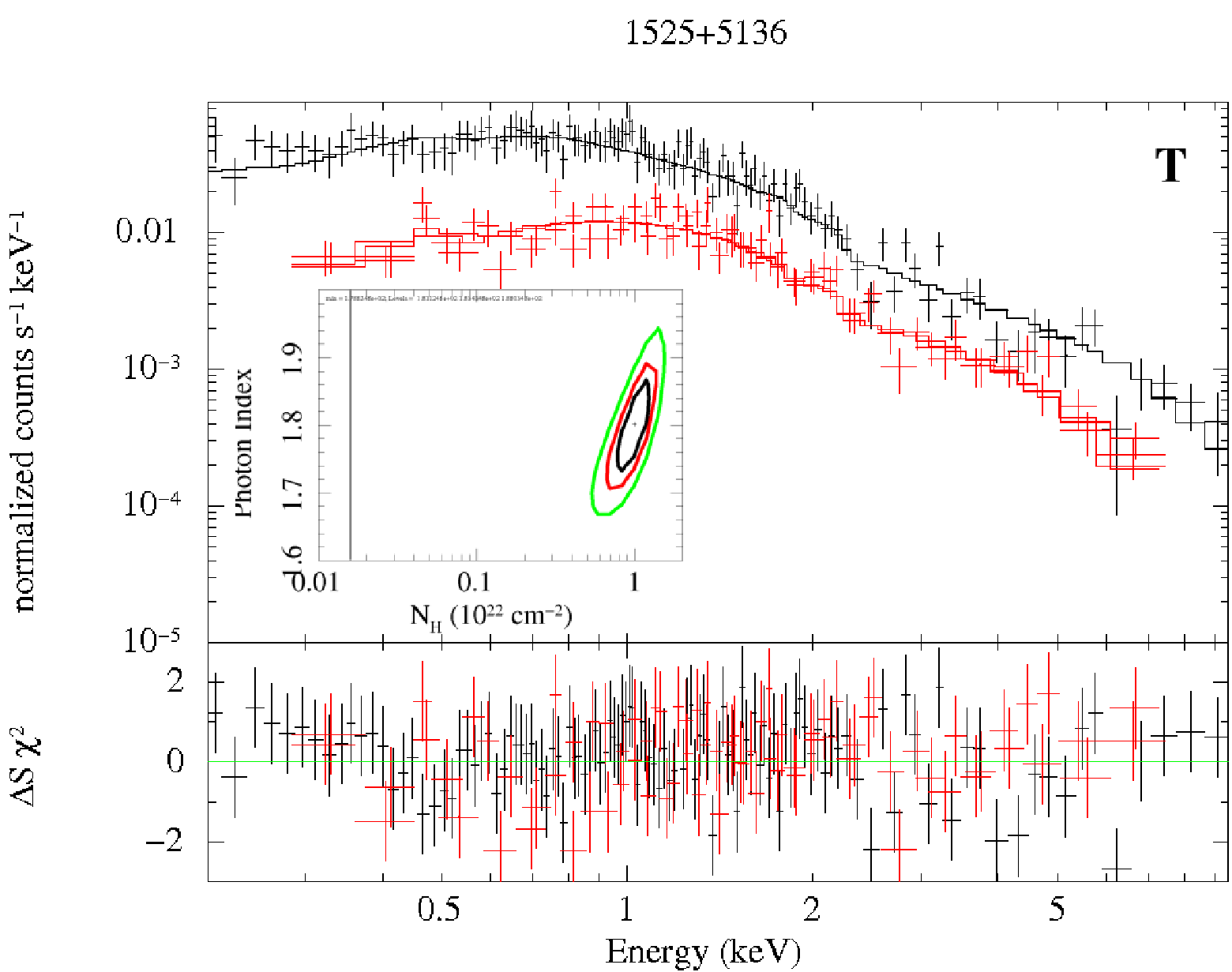}\hspace{2cm}
\includegraphics[width=8cm]{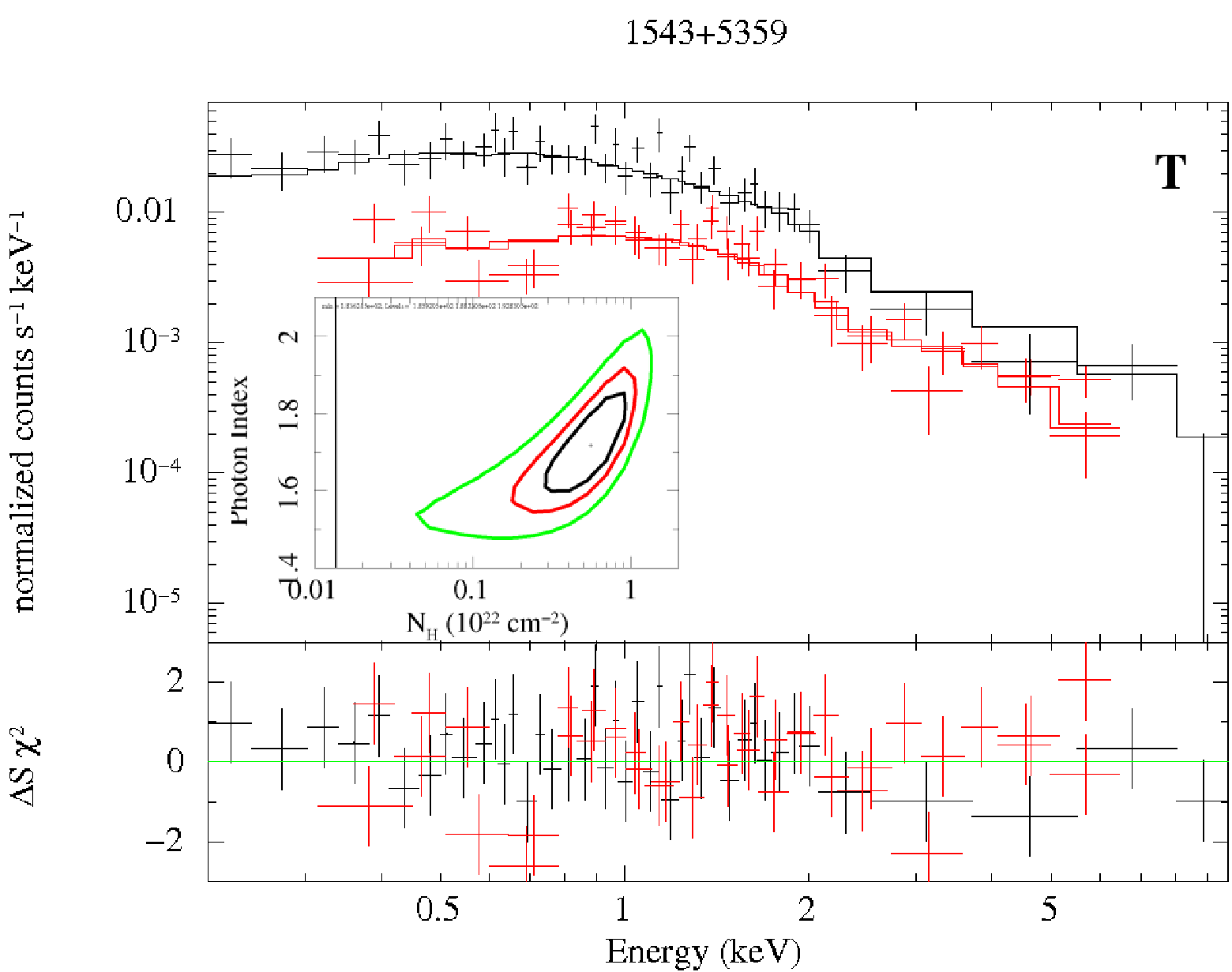}}\\%
\caption{ -- Continued}%
\end{figure*}
\end{appendix}
\end{document}